\documentstyle[]{l-aa}
\input epsf
          
\def\lsim{\lower.4ex\hbox{$\;\buildrel <\over{\scriptstyle\sim}\;$}}

\def\bib{\bibitem{}}
\newcommand{\xia}{\overline{\xi}}
\newcommand{\rhob}{\overline{\rho}}
\newcommand{\rhoa}{\overline{\rho}}
\newcommand{\gam}{\gamma}

\newcommand{\pl}{\partial}
\newcommand{\beq}{\begin{equation}}
\newcommand{\eeq}{\end{equation}}
\begin{document}
%
%
\topmargin=2.5 cm
\thesaurus{Sect.02 (12.12.1; 11.05.2; 11.09.3; 11.17.1)}
\title{The redshift evolution of Lyman-$\alpha$ absorbers}   
\author{Patrick Valageas\inst{1,}\inst{2}, Richard Schaeffer\inst{1} \and 
Joseph Silk\inst{2,}\inst{3}}
\institute{Service de Physique Th\'eorique, CEA Saclay, 91191 Gif-sur-Yvette, 
France
\and
Center for Particle Astrophysics, Department of Astronomy and Physics,
University of California, Berkeley, CA 94720-7304, USA
\and
Institut d'Astrophysique de Paris, CNRS, 98bis Boulevard Arago,
F-75014 Paris, France}
\maketitle
\markboth{Valageas, Schaeffer \& Silk: Lyman-$\alpha$ Clouds}{Valageas, 
Schaeffer \& Silk: Lyman-$\alpha$ Clouds}

\begin{abstract}

We present a model for the Lyman-$\alpha$ absorbers that treats all
objects (from the low-density forest clouds to the dense damped
systems) in a unified description. This approach is consistent with an
earlier model of galaxies (luminosity function, metallicity) but also
with the known description of the density field in the small-scale
non-linear regime. We consider two cosmological models: a critical
universe $\Omega=1$ with a CDM power-spectrum, and an open CDM
universe with $\Omega_0=0.3$, $\Lambda=0$. We reproduce the available
data on column density distribution as a function of redshift, the
value of the main new parameter, the background ionizing UV flux,
being consistent with the observed limits. This allows a
quantitatively trustable analytical description of the opacity, mass,
size, velocity dispersion and metallicity of these absorbers, over a
range of column densities spanning 10 orders of magnitude. Moreover,
together with an earlier model of galaxy formation this draws a
unified picture of the redshift evolution of structures in the
universe, from underdense clouds to massive high density galaxies,
from weak to very deep potential wells.

\end{abstract}

\keywords{large-scale structure of Universe -- Galaxies: evolution -- 
intergalactic medium -- quasars: absorption lines}

\section{Introduction}

Lyman$-\alpha$ absorption lines along the line-of-sight of remote
quasars are due to a vast class of different objects, ranging from
small subgalactic, very underdense (down to $1\%$ of the average
density of the universe!) still expanding hydrogen clouds up to the
halos of very large and very overdense ($10^4$ above the mean)
galaxies. Some have reached virial equilibrium, others are UV heated
and strongly coupled to their environment.

In this paper, we seek a quantitative analytical description of the
number of these objects as a function of their column density, and of
their internal properties. For long, the only tractable analytical
approximation to describe gravitational condensations was the
Press-Schechter (1974) approach that relates the number of condensed
objects to the early, linear, density fluctuations. This
approximation, however, is not quite suited to our purpose since in
principle it only describes objects with a density contrast that has
just reached the virialization threshold $\Delta_c \sim 200$. Our aim,
thus, is beyond the reach of this approximation. Indeed, we have
undertaken this task to take benefit of the recent progress (Valageas
and Schaeffer 1997, VS I) that allows one to describe matter
condensations of {\it any} density contrast directly in terms of the
(non-linear) density field at the epoch under consideration, the
latter in turn being related in a known way (Balian \& Schaeffer 1989,
Bernardeau 1994, Bouchet et al.1991, Colombi et al.1997) to the
initial conditions. This approach has already successfully reproduced
the luminosity function of galaxies (Valageas and Schaeffer 1998, VS
II).

\section{Multiplicity functions}

As was done for the galaxy multiplicity function in VS II, we consider
a cloud of mass M at redshift $z$ to be characterized by a density
threshold $\Delta(M,z)$ that is defined once the physical properties
of the Lyman$-\alpha$ clouds are identified. It will turn out to be
more convenient to attach to each such cloud a parameter $x$ given by
\beq
x(M,z) =  \frac{1+\Delta(M,z)}{\xia[R(M,z),z]}
\label{xnl}
\eeq
where 
\[
\xia(R) =   \int_V \frac{d^3r_1 \; d^3r_2}{V^2} \; \xi_2 
({\bf r}_1,{\bf r}_2) \;\;\;\;\; \mbox{with} \;\;\;\;\; V= \frac{4}{3}
\pi R^3
\]
is the average of the two-body correlation function $\xi_2 ({\bf
r}_1,{\bf r}_2)$ over a spherical cell of radius $R$ and provides the
measure of the (highly non-gaussian since we consider the actual
density field) fluctuations in such a cell. The function $\Delta(M,z)$
defines the external boundary $R(M,z)$ of the clouds, and is not
necessarily given by the virialization constraint
$\Delta=\Delta_c$. It will be specified in the next section. Then we
write for the multiplicity function of these clouds at a given
redshift $z$ (see VS I), in physical coordinates:
\beq
\eta(M,z) \frac{dM}{M}  = \frac{\rhob}{M} x h(x) dx     \label{etah}   
\eeq
where $\rhob$ is the mean physical density of the universe at redshift
$z$, while the mass fraction in halos of mass between $M$ and $M+dM$
is:
\beq
\mu(M,z) \frac{dM}{M} = x h(x) dx     \label{muh}
\eeq
For later convenience, it will be useful to define
\beq
\eta(x,z) \frac{dx}{x} = \frac{\rhob}{M(x,z)} x h(x) dx  \label{etax}
\eeq
as the physical number density of these halos at redshift $z$, the
mass $M(x,z)$ being specified by the choice of the density contrast
$\Delta(M,z)$. These multiplicity functions describe the number
density of objects of scale $M - M+dM$, defined by the constraint
$\Delta(M,z)$.

The scaling function $h(x)$ is a universal function that depends only
on the initial spectrum of fluctuations, and has to be taken from
numerical simulations although its qualitative behaviour is well-known
since it bears some {\it general model-independent properties} whose
origin is theoretical (Balian \& Schaeffer 1989) and which have been
well checked against observations as well as simulations (Valageas et al.1999; Colombi et al.1997):
\[
x \ll 1 \; : \; h(x) \propto x^{\omega-2} \hspace{0.3cm} ,
\hspace{0.3cm} x \gg 1 \; : \; h(x) \propto x^{\omega_s-1} \;
e^{-x/x_*}
\]
with $\omega \simeq 0.5$, $\omega_s \sim -3/2$, $x_* \sim 10$ to 20 and
\beq
\int_0^{\infty}  x \; h(x) \; dx   =  1 
\eeq
Note that $h(x)$ has to be measured only once, for a unique length
scale and epoch provided it is in the highly non-linear regime. Then
the scale-invariance of the many-body correlation functions which is
the basis of this approach (see VS I) allows one to derive the
multiplicity functions for any scale and time in the non-linear
regime, as can be seen from (\ref{etax}). The correlation function
$\xia$, that measures the non-linear fluctuations at scale $R$ can be
modelled in a way that accurately follows the numerical simulations
(see VS I for more details). We first consider the case of a critical
universe $\Omega=1$ with a CDM power-spectrum (Davis et al.1985)
normalized to $\sigma_8=0.5$. As usual we define $\sigma_8$ as the
value of the amplitude of the density fluctuations at scale $8 h^{-1}$
Mpc given by the linear theory ($\sigma^2 \neq \xia$). We use the
scaling function $h(x)$ obtained by Bouchet et al.(1991). We choose a
baryonic density parameter $\Omega_b=0.04$ and $H_0=60$ km/s/Mpc. In
fact, a power-spectrum $P(k) \propto k^n$ with $n=-2$ and the same
normalization gives very similar results. Then we study an open CDM
universe, with $\sigma_8=0.77$, $\Omega_b=0.03$ and $H_0=60$
km/s. These values are those we used previously to build a model for
galaxy formation and evolution (see VS II), so that we obtain an
overall consistent picture of the universe over a large range of
object masses and scales.

We can note that, as will be clear from our results, the properties of
the Lyman-$\alpha$ clouds we shall obtain depend mainly on the
physical model we build to describe these objects (that is how one
defines, or recognizes, such absorbers) and not much on the detailed
mass function of gravitational structures, {\it provided the above
model-independent properties are fulfilled}. The latter is of course
necessary, in order to count halos which lead to various absorption
features and to make sure that there is no internal inconsistency: one
must not count the same mass several times while keeping track of all
the mass in the universe (even if not all the mass produces
Lyman-$\alpha$ absorption lines). This is not an obvious task since
one needs to simultaneously count different types of objects defined
by several criteria. The formulation (\ref{etax}) allows one to
perform such counts in a consistent way. In our model, a given column
density can usually be produced by many different clouds and the
number of absorption lines of a given equivalent width will depend on
an integral of the mass function (with a suitable weight) over a large
range of parent halos. This will lessen the dependence on the exact
slope of $h(x)$. Hence our results are certainly very general and
robust, provided the physical picture we develop in this article is
correct. This seems confirmed by the good agreement we obtain with
observations, and this would hold as well for any power-spectrum not
too different from CDM or $P(k) \propto k^{-2}$.

\section{Properties of Lyman-$\alpha$ clouds}
\label{Properties of Lyman-alpha clouds}

\subsection{Small low-density clouds: Lyman-$\alpha$ forest}
\label{Small low-density clouds: Lyman-alpha forest}

We assume that the IGM in small low-density halos is heated by the UV
background radiation $J(\nu)$ to a temperature $T_0=3 \times 10^4$
K. As a consequence, baryonic density fluctuations are erased in
low-density regions or halos with a virial temperature $T<T_0$ over
scales of order $R_d$ with:
\beq
R_d(z) \sim \frac{1}{2} \; t_H \; C_s = \frac{1}{2} \; t_H \; \sqrt{\frac{\gam_{C_s} k T_0}{\mu m_p}}
\eeq
where $C_s$ is the sound speed, $\gam_{C_s} \sim 5/3$, $t_H$ the age of the universe and $m_p$ the proton mass. For density contrasts of order unity $R_d$ is also the  usual Jeans length. Hence, we consider a first class of objects of 
typical scale
 $R_d$, characterized by their mass $M$, and such that their
associated gravitational potential, measured by the associated
virial temperature $T_{vir}$, is lower than $T_0$ (while their actual
gas temperature, due to UV heating, is $T_0$). This defines the
average density $\rho_h$ and density contrast $\Delta(M,z)$ of these
absorbers. By definition the baryonic density is nearly constant over
their extent. These patches of matter can be regions which are still
in expansion, or even underdense areas, and small virialized
halos. 
One can note that an alternative approach to this ``smoothing''
of the baryonic density field on scales smaller than the ``Jeans length''
could be to suppress the high $k$-modes of the initial density field,
for instance Bi \& Davidsen (1997) used $P_{IGM}(k) = P_{DM}(k) \;
[1+(k/k_J)^2]^{-2}$. However, this is not possible in our present
study since we intend to model simultaneously, in a consistent way,
all Lyman-$\alpha$ absorbers. Indeed, in our model, Lyman limit and
damped Lyman systems (which we shall describe in the next sections)
are high density halos with a radius and an impact parameter which is
usually smaller than the length scale $R_d$ but with a large virial temperature. Hence, small
scale density fluctuations play a crucial role for these high column
density objects and must be properly taken into account. 

Note that although we describe here the Lyman-$\alpha$ forest lines in terms of distinct ``objects'' of size $R_d$, we would obtain the same results by directly considering the fluctuations of the density field. Indeed, the intersecting length of the line of sight with a filament is typically of order $R_d$, even if the overall length of this extended object could be much
larger. This is due to the facts that i) density fluctuations are larger on smaller scales and ii) that such a filament would not appear as a straight line but rather as a random walk. In addition, we note that we actually take into account all the matter which is distributed over the line of sight. 
These points are discussed in more details in the Appendix.
Assuming that photo-ionization equilibrium is achieved, the neutral number density $n_{HI}$ in the halo is:
\beq
n_{HI} = \frac{\alpha(T_0)}{G_1\;J_{21}} \; (1-Y) (1-Y/2) \; \left[
\frac{\Omega_b}{\Omega_0} \frac{\rho}{m_p} \right]^2
\eeq
where $\alpha(T_0) = 4.36 \; 10^{-10} T_0^{-0.75}$ s$^{-1}$ cm$^3$ is
the recombination rate, and $G_1\;J_{21}$ is the ionization rate of
neutral hydrogen (see Black 1981). As usual, $J_{21}$ is the UV
background radiation at the HI ionization threshold ($912 \AA$) in
units of $10^{-21}$ erg s$^{-1}$ Hz$^{-1}$ cm$^{-2}$ sr$^{-1}$, and
$G_1=3.2 \; 10^{-12}$ (Haardt \& Madau 1996). Here $\Omega_b/\Omega_0$
is the ratio of the baryonic density to the total density and $Y=0.26$
is the helium mass fraction. Using the parameter $x$, we write
\beq
n_{HI} = n_1 (1+\Delta)^2 =  n_1 \; x^2 \; \xia(R_d)^2
\eeq
where we define:
\beq
n_1(z) = \frac{\alpha(T_0)}{G_1\;J_{21}(z)} \; (1-Y) (1-Y/2) \; \left[
\frac{\Omega_b}{\Omega_0} \frac{\rhob(z)}{m_p} \right]^2 \; .
\label{n1(z)}
\eeq
As required by the Gunn-Peterson test, a given baryon mass fraction is
quite inefficient in producing neutral absorbing gas: $n_{HI} \ll
n_b$, for small overdensities. Since by definition the baryonic
density is roughly constant over the whole region, we neglect the
influence of the impact parameter of the line of sight which
intersects the considered patch of matter, and we assign to this
region a constant neutral column density $N_{HI}$:
\beq
N_{HI} = 2 \; n_{HI} \; 2/3 R_d = 4/3 \; n_1 \; x^2 \; \xia(R_d)^2 \;
R_d \label{NHI1x}
\eeq
(the factor $2/3$ comes simply from the average over the lines of
sight of the depth of a spherical cloud). The number of such regions
which a line of sight intersects per redshift interval is:
\beq
dn = \pi R_d^2 \; c \frac{dt}{dz} dz \; \eta(x,z) \frac{dx}{x}
\eeq
Using (\ref{etax}) we obtain:
\beq
\left( \frac{\pl^2n}{\pl lnN_{HI}\pl z} \right)_1 = \frac{3}{8} 
\frac{1}{R_d} c \frac{dt}{dz} \frac{1}{\xia(R_d)} x h(x)
\label {NHIcdd}
\eeq
Thus, the slope of the column density distribution $\pl n/\pl
lnN_{HI}$ depends on the scaling function $h(x)$, or more generally on
the multiplicity function of mass condensations. As we shall see
below, these ``objects'' are low density regions $\Delta \sim 1$, so
that $x \ll 1$. Hence we are in the domain where $h(x) \propto
x^{\omega-2}$, which together with (\ref{NHI1x}) leads to:
\beq
\left( \frac{\pl^2n}{\pl lnN_{HI} \pl z} \right)_1 \propto 
N_{HI}^{(\omega-1)/2} \label{NHI1hx}
\eeq
with $\omega \simeq 0.3$ for $n \simeq -2$ (Colombi et al.1997) where
$n$ is the slope of the initial power-spectrum $P(k) \propto k^n$
(generally speaking $\omega$ is a function of $n$). For very low
density regions, however, the multiplicity function is no longer given
by the power-law tail of $h(x)$ and shows a cutoff for $(1+\Delta)
\sim \xia^{\;-\omega/(1-\omega)} \ll 1$, corresponding to very
underdense objects surrounded by regions of even lower density. This
implies a lower cutoff for the column density distribution at
\beq
N_{HI1lower} = 4/3 \; n_1 \; R_d \; \xia(R_d)^{-2\omega/(1-\omega)}
\label{NHI1low}
\eeq
Note that although some of the clouds described in this regime are
underdense ($\Delta < 0$), they correspond to well-defined
objects. Indeed, in the non-linear regime (that is on small scales or
at late times), most of the volume of the universe is formed by very
low-density areas $\rho \sim \xia^{\;-\omega/(1-\omega)} \rhoa \ll
\rhoa$ that are even less dense than these clouds. In fact, on these
deeply non-linear scales, the average density $\rhoa$ loses the
significance it has on large scales, in the sense that it does not
define any longer a density boundary between two physically different
classes of objects.

On the high column density side, this description of the hydrogen
clouds is valid until the virial temperature $T$ reaches $T_0$. Thus
it stops at:
\beq
1+\Delta_{12} = 45 \hspace{0.8cm} \mbox{or} \hspace{0.8cm} x_{12} = \frac{45}{\xia(R_d)}
\eeq
which corresponds to an upper column density cutoff:
\beq
N_{HIupper} = 2700 \; n_1 \; R_d
\eeq
In the case of a critical universe, with a power-spectrum index $n \simeq -2$
which implies (Colombi et al.1996) $\omega \simeq 0.3$, we obtain for
a constant UV background:
\[
R_d(z) = 96 (1+z)^{-3/2} \;\mbox{kpc} \;,\; \Delta_{12} = 44 
\]
\[
x_{12} = 0.05 (1+z)^{2.7} \; , \; N_{HIlower} = 2 \; 10^6
(1+z)^{6.8} \; \mbox {cm}^{-2}
\]
\beq
N_{HIupper} = 1.5 \; 10^{13} (1+z)^{4.5} \; \mbox{cm}^{-2} \label{NHI12z}
\eeq
These regimes indeed correspond to low-density halos. The low column
density cutoff $N_{HIlower}$ is usually too low to be observed. The
upper cutoff $N_{HIupper}$ increases strongly with redshift, because
the HI density $n_{HI} \sim n_1(z)$ is proportional to the square of
the baryonic density which varies as $\rhob_b \propto (1+z)^3$. We
note that although the density contrast of these regions is not very
large ($\Delta < 44$), $\xia_{12} \simeq 1000 \; (1+z)^{-2.7}$ is
significantly larger than unity: one is actually in the deeply
non-linear regime where the density field has been greatly
distorted. Thus, regions with a small density contrast may in fact
have undergone the influence of strong non-linear effects and cannot
be modeled by linear theory or methods relying on simple corrections
to the latter. For instance, it is likely that some clouds with a
(dark matter) density close to, or even lower than, the average
density of the universe $\rhoa$ have in fact already ``collapsed'' (or
been deeply disturbed by shell-crossing) and shocked their baryonic
content. This would also affect the gas temperature.

\subsection{Galactic halos}

Massive halos with a virial temperature above $T_0$ do not see their
baryonic density profile smoothed out via the heating produced by the
UV background. Hence they constitute a second family of patches of
matter, where we have to take into account the variation of the
density with the impact parameter of the line of sight. Thus, we consider that virialized halos have a mean density
profile $\rho \propto r^{-\gam}$ with $\gam=1.8$, which is consistent with
the flat slope of the circular velocity observed in spiral galaxies,
as well as with the observed galaxy correlation function. Indeed, for deep potential wells we expect $\gam$ to correspond to the slope of the correlation function
$\xia$. In fact, massive galactic halos must
satisfy simultaneously the virialization constraint and a cooling
condition (Silk 1977, Rees \& Ostriker 1977). Similarly to the model
we developed in VS II for galaxies, small halos with a low circular
velocity verify that $\Delta=\Delta_c$ because their cooling time is
small so that their radius is given by the virialization constraint,
while massive halos have a long cooling time so that their boundary is
the cooling radius which is of the order of $R_{cool}=120$ kpc (this
simply means that we associate Lyman-$\alpha$ clouds with galaxies and
not with clusters of galaxies). This defines the functions
$\Delta(M,z)$, or $R(M,z)$, which we introduced in Sect.2.

We use for the gas temperature $T_{gas}$ of these halos the prescription:
\beq
T_{gas} = \mbox{Min} \left( \; T \; , \; 2 \; 10^6 K \; \right)
\eeq
This means that below $2 \; 10^6$ K the temperature of a large part of
the gas which can cool still remains of the order of the value it
reaches through shock-heating while for higher temperatures cooling is
so efficient it does not get much higher. This also ensures that we
recover the temperature range obtained in numerical simulations
(e.g. Miralda-Escude et al.1996). Note that the threshold $2 \; 10^6$
K also corresponds to the virial temperature where shock-heating due
to supernovae stops playing a dominant role (see VS II). However,
removing this upper cutoff for $T_{gas}$ leads to nearly identical
results which shows that our model is not very sensitive to its exact
value. Indeed, even at $z \sim 0$ halos with a high virial temperature
$T > 2 \; 10^6$ K are rather rare so that most absorption lines come
from shallower potential wells. We also model crudely the collapse of
baryons within the dark matter halos by assuming that the gas is
distributed within the potential well over a radius smaller by a
factor $\lambda < 1$ than the dark matter radius with the same density
profile $\rho \propto r^{-\gam}$:
\beq
R \rightarrow \lambda \; R \hspace{2cm} (1+\Delta) \rightarrow
\lambda^{-3} \; (1+\Delta)
\eeq
We use $\lambda = 0.4$ for deep potential wells with $T > 5 \; 10^4$ K and
a continuous transition to $\lambda = 1$ at $T = T_0$ since for
shallow objects $T < T_0$ there is no collapse (the gas is
photo-heated to a temperature larger than the virial temperature).

The neutral number density at the radius $r$ within such a halo of
external radius $R$ and density contrast $\Delta$ is:
\beq
n_{HI} = n_0 (1+\Delta)^2 \left( \frac{r}{R} \right)^{-2\gam}
\label{nHI2}
\eeq
where we defined $n_0(z)$ by:
\[
n_0(z) = \frac{\alpha(T_{gas})}{G_1\;J_{21}(z)} \; (1-Y) (1-Y/2) \;
\left[ \left( 1-\frac{\gam}{3} \right) \frac{\Omega_b}{\Omega_0}
\frac{\rhob(z)}{m_p} \right]^2 
\]
The column density $N_{HI}$ along the line of sight which intersects a
cloud of radius $R$ at impact parameter $b$ is:
\beq
N_{HI} = 2 n_0 R (1+\Delta)^2 \left( \frac{b}{R} \right)^{1-2\gam}
\int_0^{\sqrt {(R/b)^2-1}} \frac{du}{(1+u^2)^{\gam}}
\eeq
In order to simplify the calculations, we consider two limits in the
previous relation: $b \ll R$ and $b \rightarrow R$. In the case of a
small impact parameter we write:
\beq
N_{HI} \simeq 2 n_0 R (1+\Delta)^2 \left( \frac{b}{R}
\right)^{1-2\gam} I_{\infty} \; , \label{b2>}
\eeq
where we noted $I_{\infty} = \int_0^{\infty} (1+u^2)^{-\gam} du$. The
number of halos of this kind which a line of sight intersects at the
impact parameter $b$ per redshift interval is:
\beq
dn = 2 \pi b db \; c \frac{dt}{dz} dz \; \eta(x,z) \frac{dx}{x} 
\eeq
Thus we obtain for the contribution due to small impact parameters,
\beq
\begin{array}{l} {\displaystyle  \left( \frac{\pl^3n}{\pl lnN_{HI} 
\pl lnx \pl z} \right)_{2>} = \frac{3}{2(2\gam-1)} \frac{1}{R} c 
\frac{dt}{dz} }  \\  \\  {\displaystyle  \hspace{0.3cm} \times \; 
(1+\Delta)^{(5-2\gam)/(2\gam-1)} \; \left( \frac{N_{HI}}{2 n_0 R I_{\infty}}
 \right)^{-2/(2\gam-1)} x^2 h(x)  }
\end{array}
\label{dn2>}
\eeq
Here the index 2 refers to the fact that these clouds form our second
population of objects (the Lyman-$\alpha$ forest described in the
previous section is our first population) while $>$ corresponds to the
small impact parameter regime (hence high $N_{HI}$ for a given
cloud). As was noticed by Rees (1986), in such a model the density
profile of virialized halos $\rho \propto r^{-\gam}$ governs the slope
of the distribution of column densities:
\beq
\left( \frac{\pl^2n}{\pl lnN_{HI} \pl z} \right)_{2>} 
\propto N_{HI}^{-2/(2\gam-1)}       \label{slope2}
\eeq
through the dependence of the column density produced by a given halo
on the impact parameter $b$. This is consistent with observations for
$\gam \simeq 2$, which is indeed the case (for $\gam=1.8$ we have a
slope of $-0.77$). However, this power-law is only valid over a
limited range in $N_{HI}$ (which translates into the dependence on
$N_{HI}$ of the boundaries of the domain of integration in
$x$). Indeed, for a given halo, this regime only applies for impact
parameters much smaller than the halo radius (so that (\ref{b2>}) is
valid) but larger than the critical radius $R_n$ where self-shielding
becomes important and hydrogen is mainly neutral. Moreover, as we
explained in the previous section, small halos cannot be described in
this manner as the UV background smooths their density profile which
introduces the additional length scale $R_d$.

In a similar fashion, when the impact parameter is very close to the
radius of the halo, we write:
\beq
b = \left[ 1 - \frac{1}{2} \left( \frac{N_{HI}}{2 n_0 R} \right)^2
(1+\Delta)^{-4} \right] R \label{b2<}
\eeq
Note that by doing so we disregard the mass which is outside of the
considered halo. The number of halos along the line of sight is now:
\beq
\begin{array}{l}
{\displaystyle \left( \frac{\pl^3n}{\pl lnN_{HI} \pl lnx \pl z}
\right)_{2<} = \frac{3}{2} \frac{1}{R} c \frac{dt}{dz} (1+\Delta)^{-5}
} \\  \\ {\displaystyle \hspace{4cm} \times \; \left(
\frac{N_{HI}}{2 n_0 R} \right)^2 x^2 h(x) }  
\end{array}
\label{dn2<}
\eeq
Here the index $<$ refers to the large impact parameter regime. Thus,
we obtain a completely different power-law for the column density
distribution $\pl n/\pl lnN_{NHI} \propto N_{HI}^2$, which does not
depend on the halo density profile and is only due to geometrical
effects. In fact, its precise form is not important and it mainly
plays the role of a cutoff in the column density distribution: this
simply means that a given cloud mainly produces column densities
larger than a characteristic value. For a fixed column density
$N_{HI}$ we choose the transition between both regimes (small and
large impact parameter) as the point where the numbers of halos
$\pl^3n/\pl lnN_{HI} \pl lnx \pl z$ given by both approximations are
equal. In term of the variable $x$ it corresponds to
\beq
x_2 : \;\;\;\;\;\; 1+\Delta = (2\gam-1)^{(2\gam-1)/(8\gam)}
I_{\infty}^{-1/(4\gam)} \left( \frac{N_{HI}}{2 n_0 R} \right)^{1/2}
\eeq
with
\[
\left\{ \begin{array}{lll} (1+\Delta)= \lambda^{-3} \; (1+\Delta_c)  & \;\;
\mbox{if} \;\; & x<(1+\Delta_c)/\xia(R_{cool}) \\ & & \\ R= \lambda \; 
R_{cool} & \;\; \mbox{if} \;\; & x>(1+\Delta_c)/\xia(R_{cool}) \end{array} 
\right.
\]
As we can see in (\ref{b2>}) and (\ref{b2<}) it means $b \sim R$ as it
should: in fact one could simply use the formulae for $b \ll R$ up to
$b = R$ and stop there.

\subsection{Neutral cores}

Because of self-shielding, the deep cores of massive halos are not
ionized: all photons of the external UV background are absorbed by the
outer shells of the halo. Using (\ref{nHI2}), we define the optical
depth $\tau$ at the distance $l$ from the center of the halo (which is
assumed to be spherical) by:
\begin{eqnarray}
\tau & = & \int_l^R \sigma_{pi} n_0 (1+\Delta)^2 \left( 
\frac{r}{R} \right)^{-2\gam} dr  \nonumber \\ \tau & \simeq & 
\frac{1}{2\gam-1} \sigma_{pi} n_0 R (1+\Delta)^2 \left( 
\frac{l}{R} \right)^{1-2\gam}  \hspace{0.5cm} \mbox{if $\; l \ll R$}
\end{eqnarray}
where $\sigma_{pi}$ is the photo-ionization cross-section. Thus, for
each halo we define a ``neutral radius'' $R_n$, where $\tau=1$, which
determines the extension of the neutral core:
\beq
R_n = R \left( \frac{\sigma_{pi} n_0 R (1+\Delta)^2}{2\gam-1}
\right)^{1/(2\gam-1)} \; .  \label{Rn}
\eeq
Within $R_n$ all the hydrogen is neutral. Hence $n_{HI}$ is
proportional to the baryonic density (and no longer to its square),
and we obtain:
\beq
n_{HI} = n_2 (1+\Delta) \left( \frac{r}{R} \right)^{-\gam} \; ,
\label{nHI2n}
\eeq
with
\beq
n_2(z) = (1-\gam/3) (1-Y) \frac{\Omega_b}{\Omega_0} \frac{\rhob}{m_p}
\; . \label{n2(z)}
\eeq
Then, we proceed exactly as we did for the ionized part of the halos
in the previous sections. We again divide this population in two
sub-classes according to the value of the impact parameter with
respect to $R_n$. For small impact parameters we have:
\begin{eqnarray}
\lefteqn{\displaystyle \left( \frac{\pl^3n}{\pl lnN_{HI} \pl lnx 
\pl z} \right)_{2n>} = \frac{3}{2(\gam-1)} \frac{1}{R} c \frac{dt}{dz} 
(1+\Delta)^{(3-\gam)/(\gam-1)} } \nonumber \\ & & {\displaystyle
\hspace{1.5cm} \times \; \left( \frac{N_{HI}}{2 n_2 R J_{\infty}}  
\right)^{-2/(\gam-1)} x^2 h(x) }
\label{dn2n>}
\end{eqnarray}
where we defined $J_{\infty} = \int_0^{\infty} (1+u^2)^{-\gam/2}
du$. Here the $n$ in the index $2n>$ refers to the fact that we
consider the neutral cores within the population (2) halos. As was the
case in the previous sections for the ionized shells, the column
density distribution we obtain is a power-law (within a limited range)
with a slope determined by the halo density profile:
\beq
\left( \frac{\pl^2n}{\pl lnN_{HI} \pl z} \right)_{2n>} \propto 
N_{HI}^{-2/(\gam-1)} \label{slope3}
\eeq
The coefficient $2\gam$ is changed to $\gam$ because within deep
neutral cores the HI density is proportional to the baryonic density
$n_{HI} \propto \rho_b$ while in the ionized shells it is proportional
to its square $n_{HI} \propto \rho_b^2$ in the regime of
photo-ionization equilibrium. This leads to a slope steeper than found
previously: for $\gamma=1.8$ we get $-2.5$ instead of $-0.77$. When
the impact parameter is very close to $R_n$ we obtain:
\begin{eqnarray}
\lefteqn{\displaystyle   \left( \frac{\pl^3n}{\pl lnN_{HI} \pl lnx \pl z} 
\right)_{2n<} = \frac{3}{2} \frac{1}{R} c \frac{dt}{dz} 
(1+\Delta)^{(3-2\gam)/(2\gam-1)} } \nonumber \\ & & {\displaystyle
\hspace{1cm} \times \; \left( \frac{\sigma_{pi} n_0 R}{2\gam-1} 
\right)^{2\gam/(2\gam-1)} \left( \frac{N_{HI}}{2 n_2 R} \right)^2 x^2 h(x)  }
\end{eqnarray}
The transition between both regimes (small and large impact parameter)
corresponds to:
\beq
x_n : \;\;\;\; 1+\Delta \simeq \left( \frac{\sigma_{pi} n_0
R}{2\gam-1} \right)^{\gam-1} \left( \frac{N_{HI}}{2 n_2 R}
\right)^{2\gam-1}
\eeq
which occurs for $b \sim R_n$.

Finally, we have to take care of the transition between ionized
envelopes: regime (2), and neutral cores: regime $(2n)$. This
corresponds to $b=R_n$, and using (\ref{b2>}) to $N_{HI} = N_{HIn}$
with
\beq
N_{HIn} = 2 (2\gam-1) I_{\infty} \sigma_{pi}^{-1} \simeq 2 \; 10
^{18} \; \mbox{cm}^{-2}     \label{crosspi}
\eeq
which gives indeed $\tau = \sigma_{pi} N_{HI}/2 \simeq 1$. In order to
take into account, in a crude way, the non-sphericity of clouds, we
use a smooth transition between ionized and neutral regimes. Thus, we
multiply the contribution (2) by a factor
$\exp[-(N_{HI}/N_{HIn}-1)/\epsilon]$ where $\epsilon$ is a parameter
of order unity which describes the irregularity of clouds. We shall
use $\epsilon=5$ in the subsequent calculations. For $N_{HI} <
N_{HIn}$ we do not change the value we obtained previously for the
contribution (2), since the possible non-sphericity of the cloud can
only ionize the gas, which otherwise would be neutral, by decreasing
the column density along a few directions, while a cloud which would
be ionized according to the spherical model remains ionized. In a
similar fashion, we correct the contribution $(2n)$ by a
multiplicative factor smaller than unity, equal to
$1-\exp[-(N_{HIi}/N_{HIn}-1)/\epsilon]$, where $N_{HIi}$ is the column
density which would be obtained in the regime (2), if the cloud was
considered to be ionized. It is worth noting that even with such a
large spreading factor $\epsilon$ there is little smearing out of the
distribution of the column densities. In particular it is not
sufficient to fill up the dip at $\log(N_{HI}) \simeq 19$.

\subsection{Final column density distribution}

After this description of the different regimes which are involved in
the Lyman$-\alpha$ absorption lines, we can obtain the number of
clouds of a given neutral column density which a line of sight
intersects per redshift interval. We simply have to collect all the
contributions we developed above, as a given absorption line may be
produced by different kinds of objects. The integrations over $x$ of
the various distribution functions $\pl^3n/\pl lnN_{HI} \pl lnx \pl z$
are delimited by the range of validity of the physical regimes they
represent, as seen previously. We also add a lower cutoff for the
impact parameter $b>R_c$, in order not to count the galaxy luminous
cores (this is because a line-of-sight towards a remote quasar is
never chosen to cross the luminous core of a galaxy). We shall take in
the numerical calculations $R_c(z) = 1 \; (1+z)^{-3/2}$ kpc. In
addition, the distribution functions $\pl^3n/ \pl lnN_{HI} \pl lnx \pl
z$ allow us to get the distribution of a given absorption line (a
fixed neutral column density $N_{HI}$) with mass, radius, impact
parameter, or any other cloud property. This is not the case for
clouds of the first class (1), where each column density corresponds
to a specific cloud. Finally, we need to specify the value of the UV
background, and its evolution with redshift. We treat $J_{21}(z)$ as a
free parameter (see Tab.1), which we choose so as to reproduce the
column density distribution observed at the relevant redshifts while
being consistent with the observational constraints (in addition
$J_{21}(z)$ increases with time until $z \sim 2$ and then drops at low
redshifts, as predicted by usual models of structure formation, see
for instance Haardt \& Madau 1996).  Indeed, we shall see in the next
paragraph that we obtain the right shape for the column density
distribution, and the UV flux $J_{21}$ only gives the
normalization. In fact, as is well known, the latter only depends on
the combination $\Omega_b^2/J_{21}$ (apart from an additional
temperature dependence) for Lyman forest clouds and Lyman limit
systems, which is clearly seen on the definitions of $n_0(z)$ and
$n_1(z)$, see (\ref{n1(z)}). However, this is no longer true for
damped systems, which consist of neutral cores embedded within massive
halos, and only depend on $\Omega_b$, see (\ref{n2(z)}).

\section{Numerical results, $\Omega=1$}

We first consider the cloud properties we obtain with our model in a
critical universe $\Omega=1$ with a CDM power-spectrum (Peebles 1982,
Davis et al.1985) normalized to $\sigma_8=0.5$.

\subsection{Redshift evolution of the column density distribution}

\begin{figure}[htb]

\centerline{\epsfxsize=8 cm \epsfysize=7 cm \epsfbox{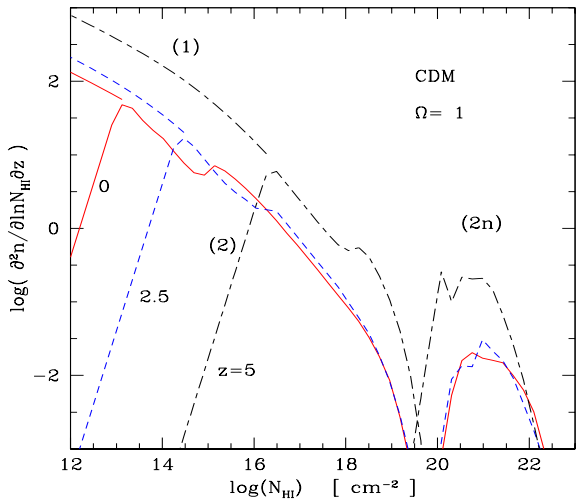}}

\caption{The contributions (1) corresponding to the Lyman-$\alpha$ 
forest (top left), (2) corresponding to the Lyman limit systems
(central part) and $(2n)$ corresponding to the damped Lyman systems
(bottom right), to the column density distribution, at $z=0$ (solid
line), $z=2.5$ (dashed line) and $z=5$ (dot-dashed line).}
\label{fdndlnNHIdz1}

\end{figure}

Fig.\ref{fdndlnNHIdz1} shows the contributions (1), (2) and $(2n)$ to
the column density distribution, at $z=0$, 2.5 and 5. We can see very
clearly on this figure the 3 classes of objects that we defined in
Sect.3.

Low column density lines ($N_{HI}< 10^{15}$ cm$^{-2}$ at $z=2.5$ and
$N_{HI}< 10^{16}$ cm$^{-2}$ at $z=5$, on the left of the picture)
correspond to low density areas and shallow potential wells
(contribution (1)) with a roughly uniform density on scales smaller
than the ``damping'' length $R_d$. The importance of this contribution
decreases at small $z$ and shifts to low column densities because as
time goes on most of the matter becomes embedded in deep potential
wells, which are described by other regimes, while the average density
of the universe declines. This translates into the redshift dependence
of the upper column density $N_{HIupper}(z)$, see (\ref{NHI12z}). These
objects exist down to quite small column densities ($N_{HI}> 10^{10}$
cm$^{-2}$ at $z=2.5$ and $N_{HI} > 10^{12}$ cm$^{-2}$ at $z=5$), that
is cloud dark matter masses as low as $10^{10} M_{\odot}$ at $z=0$ and
$10^6 M_{\odot}$ at $z=2.5$ and $z=5$. The associated absorption lines
correspond to the Lyman$-{\alpha}$ forest.

Larger column densities, up to $N_{HIn} \sim 2 \times 10^{19}$
cm$^{-2}$, are described by the regime (2). They are produced by
virialized halos, which the line of sight intersects with an impact
parameter much smaller than their radius but large enough so that the
hydrogen is highly ionized. The low column density cutoff associated
with this regime varies with redshift, as it corresponds to the
upper limit of the regime (1) we described above, while the high
column density cutoff is simply given by the ionization condition and
is of the order of $N_{HIn}$, which is the column density at which the
optical thickness is unity and whence is independent of $z$. Note that
the regime (2$<$) corresponding to the rising part on the left of
contribution (2), see eq.(\ref{dn2<}), plays no role since in this
domain most observed lines are produced by forest clouds (regime (1))
described above. Indeed, the transition (2$>$) - (2$<$) coincides with
the transition (1) - (2) as it should: we switch (almost) continuously
from one class of objects to the other one. The absorption lines
described by this regime (2) correspond to Lyman limit systems.

Finally, very high column densities (on the right of the figure),
larger than $N_{HIn}$, correspond to lines of sight which intersect
neutral cores. There is a gap in the distribution of column densities
because the fraction of neutral hydrogen switches suddenly from $\sim
10^{-3}$ in ionized shells to $1$ in neutral cores (this is linked to
the sharp change of the factor $\exp(-\tau)$ which behaves as an
exponential of the column density). As we explained in Sect.3, we can
see on the figure that the slope we obtain (for $N_{HI} \sim 10^{22}$
cm$^{-2}$) is steeper than the one we found for regime (2). These
absorption lines correspond to damped Lyman systems.

\begin{figure}[htb]

\centerline{\epsfxsize=8 cm \epsfysize=7 cm \epsfbox{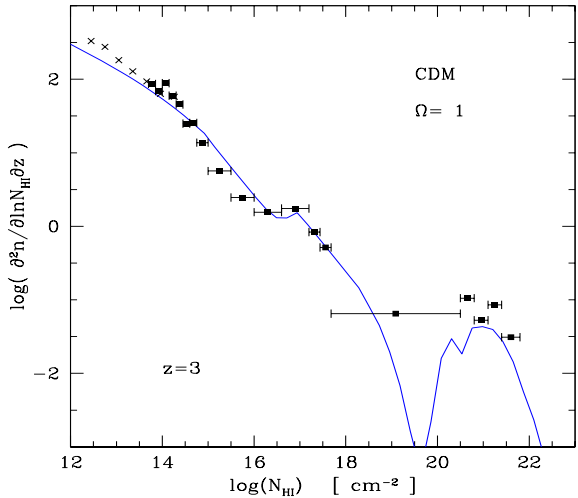}}

\caption{The total column density distribution (summed over all 
contributions: Lyman-$\alpha$ forest, Lyman-limit and damped
Lyman-$\alpha$ systems) at $z=3$. The data points are from Hu et
al.(1995) (crosses) and from Petitjean et al.(1993) (rectangles).}
\label{fdntotodNHIdz1}

\end{figure}

We show in Fig.\ref{fdntotodNHIdz1} the total column density
distribution at $z=3$ which results from the sum over all
contributions. We can see that our model agrees reasonnably well with
the observations, from $N_{HI} \sim 10^{13}$ cm$^{-2}$ up to $N_{HI}
\sim 10^{22}$ cm$^{-2}$. In particular, we recover the under-abundance
of lines at intermediate column densities $N_{HI} \sim 10^{15} -
10^{16}$ cm$^{-2}$. This feature is also quite clear on
Fig.\ref{fdndlnNHIdz1}. It corresponds to the transition between
populations (1) and (2). The gas embedded within galactic halos can
cool and collapse (for deep potential wells $T > T_0$) which increases
the column density along the line-of-sight (the neutral number density
scales as the square of the baryonic density). As a consequence,
clouds which would have produced column densities somewhat larger than
$N_{HIupper}(z)$ actually lead to deeper absorption features. This
produces an under-abundance of objects around $N_{HI} \sim 10^{15}$
cm$^{-2}$ at $z=2.5$.

\begin{figure}[htb]

\centerline{\epsfxsize=8 cm \epsfysize=7 cm \epsfbox{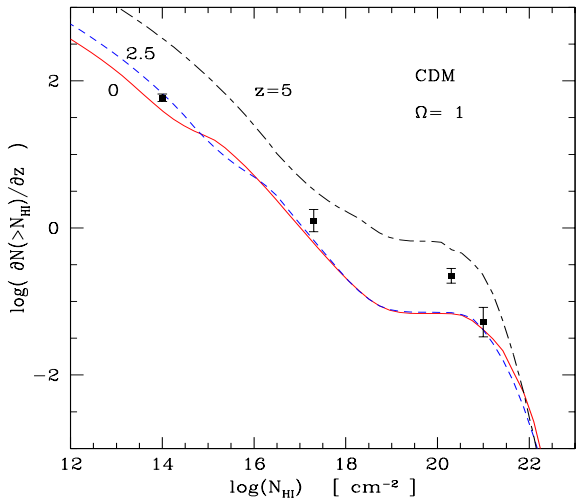}}

\caption{The total column density cumulative distribution at $z=0$ 
(solid line), $z=2.5$ (dotted line) and $z=5$ (dot-dashed line). The
data are from Bechtold (1994) (Forest at $z=2.5$), Lanzetta et al.(1995)
(Lyman limit systems at $z=2.6$) and Wolfe et al.(1995) (damped Lyman
systems at $z=2.25$).}
\label{fdnNHIdz1}

\end{figure}

Fig.\ref{fdnNHIdz1} shows the total (summed over all contributions)
column density cumulative distribution at $z=0, 2.5$ and $5$. The flat
part for $10^{19} < N_{HI} < 10^{20}$ cm$^{-2}$ corresponds to the gap
in Fig.\ref{fdndlnNHIdz1}. This is also seen in numerical simulations
(Katz et al.1996). We can see that the slope of the distribution
function we obtain is again consistent with observations.

\begin{figure}[htb]

\centerline{\epsfxsize=8 cm \epsfysize=7 cm \epsfbox{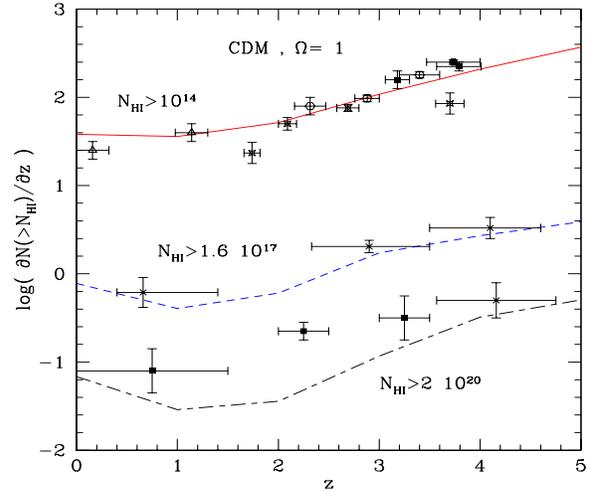}}

\caption{The evolution with redshift of the total column density 
cumulative distribution for $N_{HI} > 10^{14}$ cm$^{-2}$ (upper
curve), $N_{HI} > 1.6 \; 10^{17}$ cm$^{-2}$ (intermediary curve) and
$N_{HI} > 2 \; 10^{20}$ cm$^{-2}$ (lower curve). The data points are
from Bechtold (1994) (stars), Bahcall et al.(1996) (triangles), Giallongo
et al.(1996) (filled squares), Kim et al.(1997) (open circles) and Lu et
al.(1996) (filled circle) for $N_{HI} > 10^{14}$ cm$^{-2}$; from
Storrie-Lombardi et al.(1994) (crosses) for $N_{HI} > 1.6 \;
10^{17}$ cm$^{-2}$; and from Wolfe et al.(1995) (filled squares) and
Storrie-Lombardi et al.(1995) (cross) for $N_{HI} > 2 \;
10^{20}$ cm$^{-2}$.}
\label{fdndz1}

\end{figure}

\begin{table}
\begin{center}
\caption{Redshift evolution of the background UV flux $J_{21}(z)$ used 
in this article for both cosmologies (critical and open universe).}

\begin{tabular}{ccccccc}\hline

 &  $z=$0  &  1  &  2  &  3  &  4  &  5  \\
\hline \hline
\\

$\Omega=$1 & $J_{21} =$ 0.05 & 0.4 & 0.7 & 0.4 & 0.2 & 0.1 \\

$\Omega_0=$0.3 & 0.05 & 0.5 & 0.8 & 0.4 & 0.2 & 0.1 \\

\end{tabular}
\end{center}
\label{table1}
\end{table}

The evolution with redshift of the total column density cumulative
distribution is shown in Fig.\ref{fdndz1} for $N_{HI}>10^{14}$
cm$^{-2}$, $N_{HI}>1.6 \; 10^{17}$ cm$^{-2}$ and $N_{HI}>2 \; 10^{20}$
cm$^{-2}$. We can see in the figure that we can fit the data
simultaneously for the 3 types of absorption lines which are
representative of 3 different classes of objects: Lyman$-\alpha$
forest clouds, Lyman limit systems and damped Lyman systems, which are
indeed described in our model by 3 different regimes. This builds our
confidence in the validity of this model. Note that we could increase the number of damped systems by using a slightly larger $\sigma_8$ which does not significantly modify our results for limit and forest objects, as shown below in Sect.\ref{Influence of various parameters}. With $\Omega_b=0.04$, the
values of the UV flux we use are shown in Tab.1. They are consistent
with observations: Giallongo et al.(1996) find $J_{21} = 0.5 \pm 0.1$
for $1.7<z<4.1$, while Cooke et al.(1997) get $J_{21} =
1^{+0.5}_{-0.3}$ for $2<z<4.5$, with no evidence for any redshift
evolution within these intervals. At low redshifts $z \leq 1$ the UV
flux shows a sharp drop (at $z=0$ Vogel et al.1995 find $J_{21} <
0.07$ while Donahue et al.1995 get $J_{21} < 0.033$). Moreover, the
overall redshift dependance of $J_{21}(z)$ shown in Tab.1 is similar
to the prediction of the usual models of galaxy formation from
hierarchical scenarios where the radiation comes from stars and
quasars (e.g. Haardt \& Madau 1996). We can see in Fig.\ref{fdndz1}
that we recover the observed break in the redshift evolution of the
number density of lines at $z \sim 1.5$ (e.g. Jannuzi 1998). Indeed,
at lower $z$ the number of lines with $N_{HI} > 10^{14}$ cm$^{-2}$
remains constant while it slightly increases for Lyman-limit and
damped systems. This break in the continuous decline with time of the
number of forest lines due to structure formation which builds
increasingly deep potential wells is produced by the sudden drop of
the UV flux at low $z$. At low redshifts $z \sim 0$ we may
overestimate the number of lines with $N_{HI} > 1.6 \; 10^{17}$
cm$^{-2}$ and $N_{HI} > 2 \; 10^{20}$ cm$^{-2}$ which come from
galactic halos since we did not take into account star-formation which
consumes (and may eject) some of the gas.

We can note that we manage to get a satisfactory agreement
with observations for the column density distribution of
Lyman-$\alpha$ clouds, while using $\Omega_b=0.04, \; H_0 =$ 60 km/s,
$\sigma_8=0.5$ and $J_{21}(z=2.5)=0.7$. On the other hand, numerical
simulations need $\Omega_b>0.05$ for $J_{21}(z=2.5)>0.2$,
$\sigma_8=0.79$, $\Omega=0.4$, $\lambda=0.6$, and $H_0=$ 65 km/s
(Miralda-Escude et al.1996) or $\Omega_b>0.05$ for
$J_{21}(z=2.5)>0.1$, $\sigma_8=0.7$, $\Omega=1$ and $H_0=$ 50 km/s
(Katz et al.1996). These latter results would mean that it is
difficult to satisfy the nucleosynthesis bounds with the observational
estimates of $J_{21}$. The fact that we do not encounter such a
serious problem here is rather encouraging and suggests, as we shall see below, that due to the very extended range of clouds (from weak
potential wells to very deep halos) which contribute to a given
$N_{HI}$ it is difficult to take into account properly all
contributions (especially from the shallower clouds) in a simulation,
which should thus usually underestimate the column density
distribution function.

\subsection{Influence of various parameters}
\label{Influence of various parameters}

\begin{figure}[htb]

\centerline{\epsfxsize=8 cm \epsfysize=7 cm \epsfbox{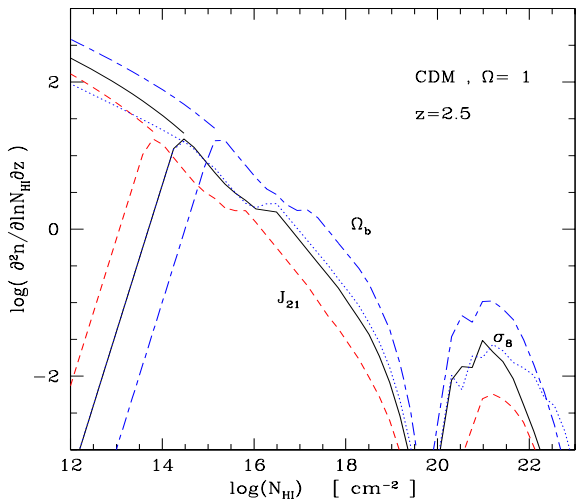}}

\caption{The column density distribution as in Fig.\ref{fdndlnNHIdz1} 
at $z=2.5$ (solid line), with $J_{21}=3$ instead of 0.7 (dashed line),
with $\Omega_b=0.1$ instead of $0.04$ (dot-dashed line), or with
$\sigma_8=1.3$ instead of 0.5 (dotted line), at the same redshift.}
\label{fdndlnNHIdzJOSig1}

\end{figure}

As we explained in Sect.3, for all regimes the slope of the column
density distribution only depends on the shape of the initial
power-spectrum through $\omega$ and $\gam$ (we take $\gam=1.8$ but it
could slightly depend on the slope $n$ of $P(k)$) as we can see in
(\ref{NHI1hx}), (\ref{slope2}) and (\ref{slope3}). This constrains $n$
to be close to $-2$ hence both a CDM-like power-spectrum (since the
local slope on the scales of interest is indeed close to $-2$) and a
power-law power-spectrum with $n \simeq -2$ give satisfactory
results. However, the normalization of the column density distribution
depends on the UV flux $J_{21}$, as we noticed above, and on
cosmological parameters. Indeed, we can see from Sect.3 that we have:
\[
\left\{ \begin{array}{l} {\displaystyle  \left( \frac{\pl^2n}{\pl 
lnN_{HI} \pl z} \right)_1 \propto \left( \frac{N_{HI} R_d}{n_1}
\right)^{-(1-\omega)/2} \; [ \xia(R_d) R_d ]^{-\omega} h^{-1} } \\ \\
{\displaystyle \left( \frac{\pl^2n}{\pl lnN_{HI} \pl z} \right)_{2>}
\propto \left( \frac{N_{HI}}{n_1} \right)^{-2/(2\gam-1)} \; h^{-1} }
\\ \\ {\displaystyle \left( \frac{\pl^2n}{\pl lnN_{HI} \pl z}
\right)_{2n>} \propto \left( \frac{N_{HI}}{n_2} \right)^{-2/(\gam-1)}
\; h^{-1} } \end{array} \right.  \label{sys}
\]
with
\beq
n_1 \propto \frac{T_0^{-0.75} (\Omega_b h^2)^2}{J_{21}} \hspace{0.6cm}
\mbox{and} \hspace{0.6cm} n_2 \propto \Omega_b h^2 \label{JsigOm}
\eeq
where we neglected the influence of the boundaries of the integrations
over $x$. Fig.\ref{fdndlnNHIdzJOSig1} shows the influence of various
parameters on the column density distribution. If we increase the
normalization $\sigma_8$ of the power-spectrum hence $\xia$ (dotted
line in the figure), the number of Lyman-$\alpha$ forest lines
decreases: clustering is more important (one is deeper into the
non-linear regime) so that a larger fraction of the mass of the
universe is within high density virialized halos which produce Lyman
limit and damped systems. Simultaneously these two latter
contributions increase, but this is not shown in (\ref{JsigOm})
because we neglected the variation of the boundary in $x$. This
increase is most important for highest column densities (massive
damped systems) since these lines come from the rare high density
halos which are in the exponential cutoff of the mass functions and
whose abundance is very sensitive to the normalization of the
power-spectrum (it enters as an exponential factor). On the other hand
the number of Lyman-limit lines is nearly invariant because each of
them is drawn from a large population of possible parent halos:
increasing $\sigma_8$ means there are fewer low mass halos but more
large mass halos and both effects nearly cancel each other. If the
baryonic density parameter $\Omega_b$ gets higher (dot-dashed line)
all contributions increase since there is more hydrogen available but
the magnitude of this effect is not the same for all lines. Finally,
if we increase $J_{21}$ the number of lines will decrease in all
regimes since the neutral fraction of hydrogen is lower (dashed
line). Once again, this is not seen in (\ref{JsigOm}) for damped
systems because it appears through the change of the boundaries over
$x$: the deep cores of virialized halos are not influenced by a small
increase of $J_{21}$, which only destroys the outer neutral
shells. Indeed we can check in Fig.\ref{fdndlnNHIdzJOSig1} that the
number of very large column density lines does not change. We note
that a change of $\Omega_b$ or $J_{21}$ simply leads to a horizontal
translation of the column density distribution for forest and Lyman
limit absorbers. Indeed, the number and properties (for the dark
matter) of these halos do not change, but a given region will produce
a column density which evolves with the baryonic fraction ($\Omega_b$)
and the neutral hydrogen fraction ($J_{21}$). This is no longer true
for damped systems where a change of $\Omega_b$ or $J_{21}$ usually
introduces new halos or removes the smallest ones. Moreover, we can
see that the location of the gap between Lyman limit and damped
systems is constant since it only depends on the photo-ionization
cross-section, see (\ref{crosspi}).

Thus, we see that the normalization of the column density distribution
does not depend only on the usual parameter $n_1$, and that a change of a
given variable ($\Omega_b, J_{21}, T_0$ or $\sigma_8$) usually
influences the three classes of objects described in Sect.3 in a
different way. The dependence on $\sigma_8$ is very small for all
values of interest except for the largest damped systems. This means
that a power-spectrum normalized to the COBE data would also lead to a
reasonable agreement with observations of Lyman-$\alpha$ clouds. The
influence of $\Omega_b$ and $J_{21}$ is stronger, but it is degenerate
for forest and Lyman limit lines through the combination $n_1$. Thus,
for a given normalization of the power-spectrum, one could first
derive $\Omega_b$ from observations of large damped systems, and then
obtain $J_{21}$ from Lyman limit or forest absorbers, in order to
match the data. However, we must emphasize that we have only one
important ``free'' parameter in our model: $J_{21}$ (which must also
be consistent with observations), since $P(k)$ and $\Omega_b$ are
taken from consistent models of galaxies (VS II) and clusters which were already fully constrained
by other sets of observations. This implies for instance that the
abundance of the largest damped lines is given by these previous
models. Indeed, as we explained above, these lines correspond simply
to deep neutral shells of galactic halos, while the forest absorbers
are new low-density objects (predicted by the same description of the
non-linear density field, but in a different density regime)

\subsection{Mass and circular velocity associated to different clouds}

\begin{figure}[htb]

\centerline{\epsfxsize=8 cm \epsfysize=7 cm \epsfbox{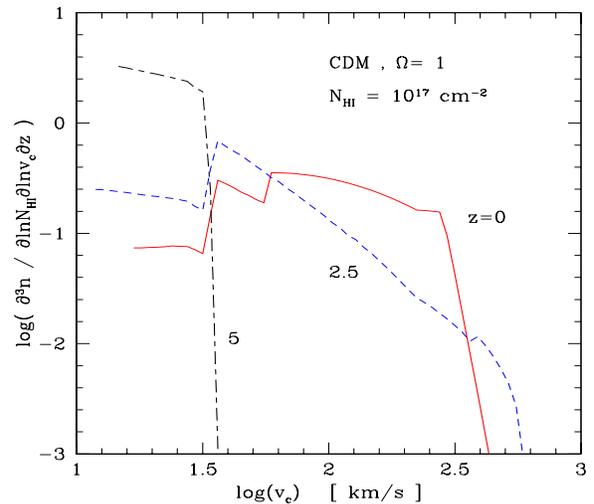}}

\caption{Evolution with redshift of the halo rotational velocity 
distribution function, for $N_{HI}=10^{17}$ cm$^{-2}$, at $z=0$ (solid
line), $z=2.5$ (dashed line) and $z=5$ (dot-dashed line).}
\label{fdndlnNHIdlnvcdz1}

\end{figure}

Our model yields the evolution with redshift of the halo rotational
velocity (or mass, radius,..) distribution function, for a fixed
column density. We define the halo circular velocity $v_c$ as:
\beq
v_c^2 = \frac{{\cal G} M}{R} = \frac{4\pi}{3} {\cal G} \rhoa
(1+\Delta) R^2
\eeq
and Fig.\ref{fdndlnNHIdlnvcdz1} presents the case $N_{HI}=10^{17}$
cm$^{-2}$ at $z=0, 2.5$ and $5$. The average halo rotational velocity
(or mass) gets larger as time goes on, since gravitational clustering
builds increasingly deep and massive potential wells. The sharp high
velocity cutoff is not due to the multiplicity function of virialized
halos but to the fact that very large and massive clouds cannot
produce column densities as low as $N_{HI}=10^{17}$ cm$^{-2}$. In
other words, looking at a specific $N_{HI}$ selects a finite range of
parent halos (the contribution of larger clouds is not exactly zero
because if the line of sight intersects such a halo very close to its
external radius it can still produce a small column density, due to
the small intersection length, but this occurs with a negligible
probability as can be seen in the figure). The location of this cutoff
does not evolve much from $z=0$ to $z=2.5$ because in this range
clouds are defined by the cooling relation $R=R_{cool}$. Indeed, for a
given halo the characteristic ``minimum'' column density on a line of
sight is given by:
\beq
N_{HImin} \sim n_1 (1+\Delta)^2 R \sim \left( \frac{v_c}{R \sqrt{{\cal
G} \rhoa}} \right)^4 n_1 R
\eeq
similarly to the calculation described previously for regime (1)
clouds, see (\ref{NHI1x}). Hence, a given column density selects a
``maximum'' rotational velocity:
\beq
v_{cmax} \sim R \sqrt{{\cal G} \rhoa} \left( \frac{N_{HI}}{n_1 R}
\right)^{1/4} \propto \left( \frac{J_{21}(z) N_{HI}
R_{cool}^3}{\Omega_b^2} \right)^{1/4} \label{vcRcool}
\eeq
which evolves slowly with $z$ through $J_{21}(z)^{1/4}$. In fact, at
low $z$ this upper velocity cutoff increases with $z$ because of the
sharp decline of $J_{21}(z)$ with time. Indeed, at $z \sim 2$ in order
to observe a given column density (here $N_{HI}=10^{17}$ cm$^{-2}$)
one must look through a deeper potential well (larger mass and
density) defined by a higher $T$ and $v_c$ than at $z \sim 0$ because
$J_{21}$ is much larger. On the other hand, at $z=5$ the clouds
corresponding to $v_{cmax}$ for $N_{HI}=10^{17}$ cm$^{-2}$ are defined
by the virialization condition $\Delta=\Delta_c$, so that we get:
\beq
v_{cmax} \propto \frac{J_{21}(z)}{\Omega_b^2}
\frac{N_{HI}}{[(1+\Delta_c) \rhoa]^{3/2}} \propto J_{21}(z)
(1+z)^{-9/2} \label{vcDeltac}
\eeq
which decreases strongly at high redshift (note that we neglected the
influence of the collapse factor $\lambda$). This behaviour can be
understood very simply within the framework of the physical picture we
developed in Sect.3 for Lyman-limit clouds, based on the association
of these absorbers with galaxies. Indeed, at high redshift as time
goes on the hierarchical clustering process builds increasingly
deeper and more massive but lower density potential wells (the density
of these halos scales as $(1+\Delta_c) \rhoa$), so that the rotational
velocity attached to a given column density gets larger with time, as
in (\ref{vcDeltac}). However, after some time (typically $z<1-2$)
galaxies are no longer defined by the sole virialization constraint:
massive galaxies are determined by a cooling constraint (VS II, Silk
1977, Rees \& Ostriker 1977) which introduces a fixed length scale
$R_{cool}$. As a consequence, the highest velocity associated with a
given $N_{HI}$ does not evolve any more (except with the possible
changes of the background UV flux $J_{21}$) which leads to the regime
described by (\ref{vcRcool}). In other words, at small redshift we
identify Lyman-limit clouds with galaxies, and not with clusters of
galaxies. This leads to a qualitative change in the behaviour of some
physical properties. Thus, the ``simple'' hierarchical clustering
scenario (coupled with a proper physical picture) can produce
non-trivial histories, that is with qualitative discontinuities and
amenable to observational checks. The lower cutoff is due to the fact
that low velocity halos are described by regime (1): their density is
roughly uniform over scales of the order of the ``damping'' length $R_d$ and they cannot produce large column density absorption lines. Thus, it
corresponds to the virial temperature $T_0$ and shows (nearly) no
evolution with redshift. The small discontinuity in the curve at $z=0$
around $v_c = 60$ km/s corresponds to the transition from halos
defined by $\Delta=\Delta_c$ (virialization constraint) to
$R=R_{cool}$ (cooling constraint). The latter discontinuity thus has
no physical meaning and is simply due to our interpolation procedure
between these different regimes.

\begin{figure}[htb]

\centerline{\epsfxsize=8 cm \epsfysize=12 cm \epsfbox{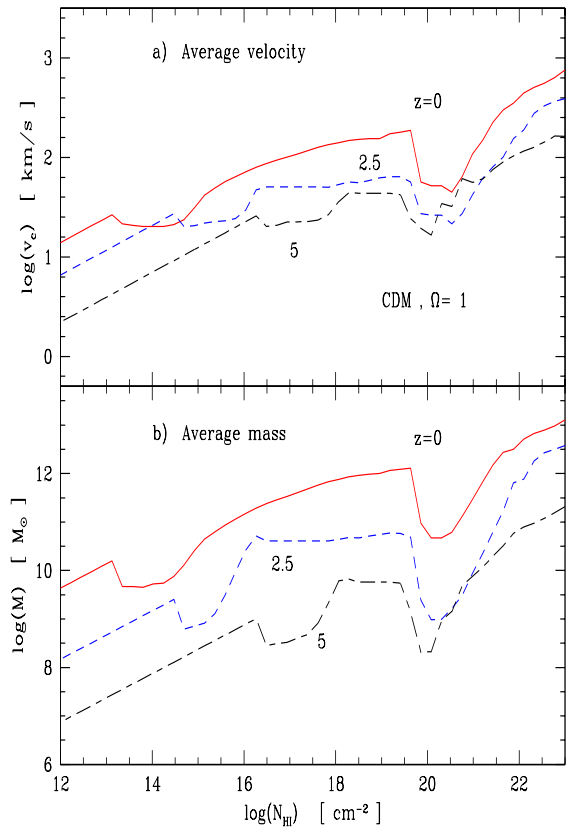}}

\caption{a) Evolution with redshift of the average rotational velocity 
of the halos associated to a given column density at $z=0$ (solid
line), $z=2.5$ (dotted line) and $z=5$ (dot-dashed line). b)
Evolution with redshift of the average mass of the halos associated to
a given column density, for the same cases.}
\label{fvcMaNHI1}

\end{figure}

\begin{figure}[htb]

\centerline{\epsfxsize=8 cm \epsfysize=12 cm \epsfbox{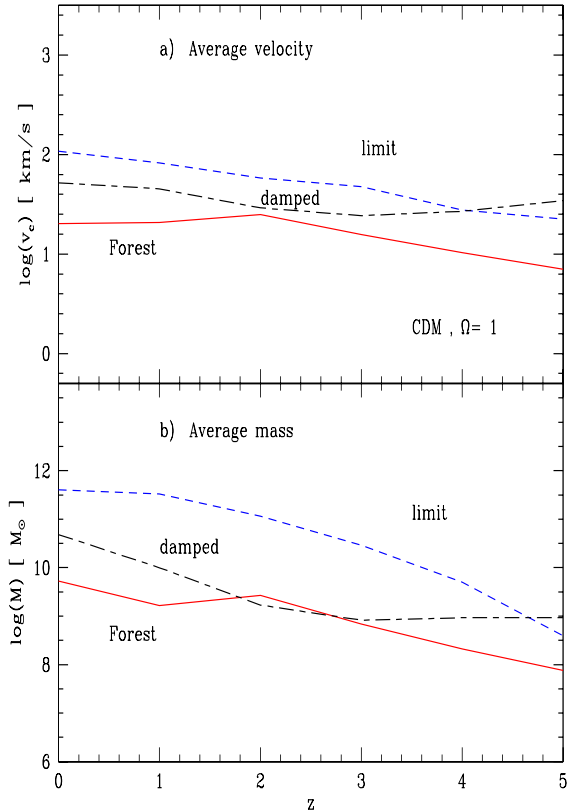}}

\caption{a) Evolution with redshift of the mean halo rotational velocity, 
for clouds defined by $N_{HI} = 10^{14}$ cm$^{-2}$ (solid line),
$N_{HI} = 10^{17}$ cm$^{-2}$ (dashed line) and $N_{HI} = 1.6 \;
10^{21}$ cm$^{-2}$ (dot-dashed line). b) Evolution with redshift of
the mean halo mass, for the same cases.}
\label{fvcMaz1}

\end{figure}

We note in Fig.\ref{fdndlnNHIdlnvcdz1} that the velocity distribution
is quite extended, and that for $z \leq 2.5$ it is not far from
uniform from 20 km/s up to 400 km/s. This is due to the fact that a
given cloud may produce many different absorption features according
to the value of the impact parameter $b$. Conversely, a given column
density $N_{HI}$ may originate with a similar probability from many
different clouds. Even at $z=0$, small velocity clouds ($v_c < 100$
km/s) make up a sizeable fraction ($\sim 40 \%$) of the total
Lyman-$\alpha$ lines with $N_{HI}= 10^{17}$ cm$^{-2}$. This could
explain the fact that simulations with insufficient resolution produce
fewer Lyman-limit systems than is required by the observations (Katz
et al.1996) since it appears that one should resolve clouds down to
$v_c \sim 20$ km/s.

Alternatively, we can consider the evolution with redshift of the mean
halo rotational velocity or dark matter mass associated with a given
column density. This is displayed in Fig.\ref{fvcMaNHI1} for $z=0,
2.5$ and $5$. The feature at low velocity and mass corresponds to the
transition between regimes (1) and (2), while the sudden change at
$N_{HI} \sim 10^{20}$ cm$^{-2}$ is due to the transition from lines of
sight intersecting ionized shells to those probing the neutral cores
of deep halos. For any column density the average velocity and mass
grow as time increases because the hierarchical clustering process
builds deeper and more massive halos. At small $N_{HI}$ in regime (1),
there is a unique correspondance between column density and velocity,
or mass, see (\ref{NHI1x}), which gives:
\beq
N_{HI} \propto (1+\Delta)^2 \propto v_c^2 \propto M^2
\eeq
since the radius $R=R_d$ is constant. For larger $N_{HI}$ there is no
longer such a unique relation, as different clouds can produce the
same column density. The mean velocity or mass first increases with
$N_{HI}$ and then reaches a plateau where it is nearly independent of
$N_{HI}$. The rising part of the curve corresponds to the fact that
larger column densities can be produced by deeper and more massive
halos (as we described above for the high cutoff of the velocity
distribution at fixed $N_{HI}$). Hence, as one looks for larger
equivalent widths one adds to the population of parent halos new more
massive and deeper clouds (while the minimum mass of the possible
halos does not change since it is given by the fixed transition to the
regime (1), one simply needs to draw lines of sight which pass closer
to the center of this small potential well). As a consequence the
average velocity (mass) increases with $N_{HI}$, until the high
velocity (mass) cutoff gets larger than the typical velocity (mass) of
objects which have collapsed at the considered redshift (this is where
the cutoff of the mass functions comes in). Then, adding a new
population of more massive halos when looking at a larger $N_{HI}$
only produces a negligible change in the mean velocity or mass because
the number of these new clouds is very small.  

Finally, there is a
transition at $N_{HI} \sim 10^{20}$ cm$^{-2}$ where the average
velocity and mass suddenly decrease to grow again later with
$N_{HI}$. Indeed, for column densities slightly larger than $10^{20}$
cm$^{-2}$ we start to probe the neutral cores of virialized halos. The
sharp change in the fraction of neutral hydrogen reflects that column
densities $N_{HI} \sim 10^{20} - 10^{21}$ cm$^{-2}$ can be produced by
small clouds which were characteristic of the lowest Lyman-limit
lines: the maximum velocity or mass allowed for a cloud to produce
such a line decreases suddenly to the value corresponding to the
transition between regimes (1) and (2), hence we recover the same mean
velocity and mass, as can be seen in the figure. Then, exactly as
occured for Lyman-limit systems, the average velocity (mass) first
increases with $N_{HI}$ and finally reaches a plateau, due to the
cutoff of the mass functions. This final velocity (mass) is larger
than the one reached in the regime (2) for Lyman-limit lines, because
the velocity (mass) distribution is more heavily weighted towards
large velocity (mass). This is due to two factors. First, the neutral
hydrogen density $n_{HI}$ is now proportional to the baryonic density
and not to its square, see (\ref{nHI2}) and (\ref{nHI2n}), which means
that for a given cloud the change with the impact parameter $b$ of the
column density is slower but also that for slightly larger clouds the
impact parameter must increase faster in order to produce the same
$N_{HI}$, which leads to a cross-section factor $b^2$ more heavily
weighted towards large clouds. Second, the neutral radius $R_n$ grows
faster than $R$, see (\ref{Rn}), which again favors massive halos. Of
course, this translates directly into the $R$ and $\Delta$ dependence
of $\pl^3n/ \pl lnN_{HI} \pl lnx \pl z$ in both regimes, see
(\ref{dn2>}) and (\ref{dn2n>}).

Fig.\ref{fvcMaz1} shows directly the redshift evolution of the average
velocity or mass for 3 types of clouds, representative of the 3
classes of objects we described previously. The main feature is that
for all clouds the mean velocity and mass decrease at higher redshift,
since only weaker and smaller potential wells were formed at earlier
times. This is not always true for the high column density $N_{HI} =
1.6 \; 10^{21}$ cm$^{-2}$ because of the transition between ionized
shells and neutral cores which introduces additional effects, as
described above. This also explains why the average mass or velocity
associated with this damped system is smaller at low $z$ than those
characteristic of a lower column density Lyman-limit system.

\subsection{Radius and impact parameter associated to various 
column densities}

\begin{figure}[htb]

\centerline{\epsfxsize=8 cm \epsfysize=7 cm \epsfbox{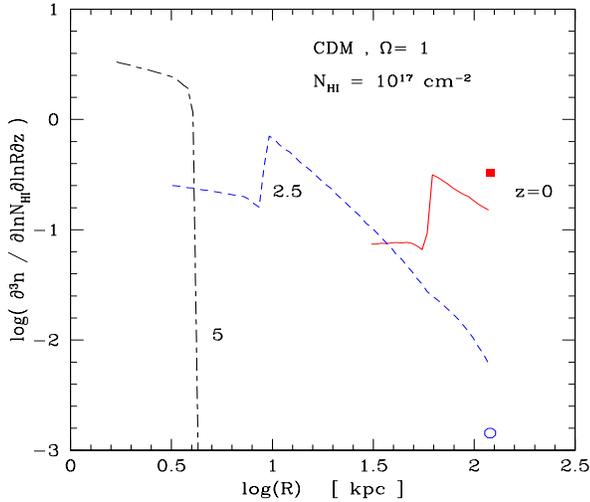}}

\caption{Evolution with redshift of the halo radius distribution 
function, for $N_{HI}=10^{17}$ cm$^{-2}$, at $z=0$ (solid line),
$z=2.5$ (dotted line) and $z=5$ (long dashed line). The filled
rectangle and the small ellipse represent the contribution of the
clouds defined by the constraint $R=R_{cool}$ at $z=0$ and $z=2.5$,
see text.}
\label{fdndlnNHIdlnRdz1}

\end{figure}

\begin{figure}[htb]

\centerline{\epsfxsize=8 cm \epsfysize=12 cm \epsfbox{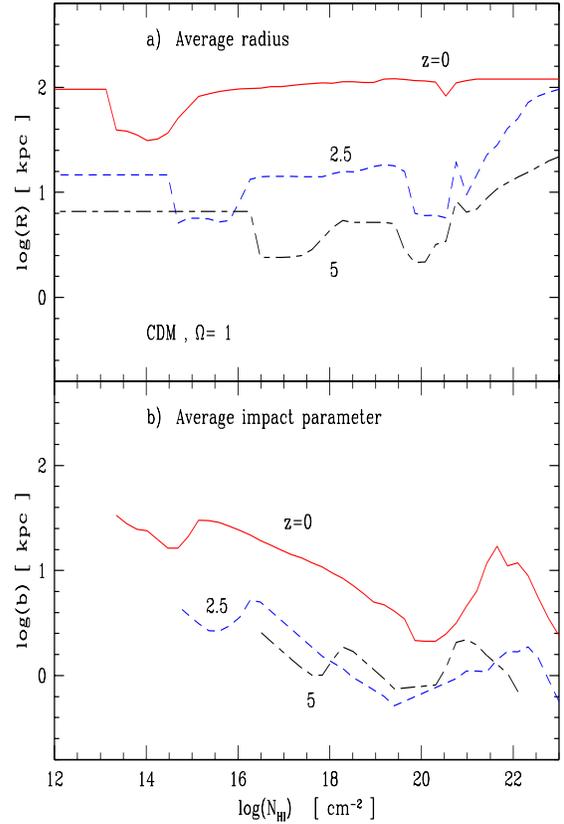}}

\caption{a) Evolution with redshift of the average radius and impact 
parameter of the halos associated to a given column density at $z=0$
(solid line), $z=2.5$ (dotted line) and $z=5$ (dot-dashed line). b)
Evolution with redshift of the average neutral fraction of the halos
associated to a given column density, for the same cases.}
\label{fRbfaNHI1}

\end{figure}

\begin{figure}[htb]

\centerline{\epsfxsize=8 cm \epsfysize=12 cm \epsfbox{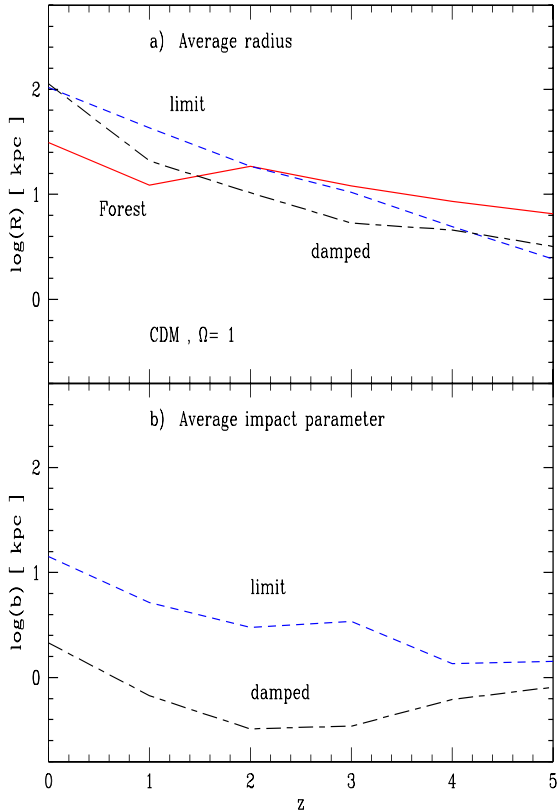}}

\caption{a) Evolution with redshift of the mean radius, for clouds 
defined by $N_{HI} = 10^{14}$ cm$^{-2}$ (solid line), $N_{HI} =
10^{17}$ cm$^{-2}$ (dashed line) and $N_{HI} = 1.6 \; 10^{21}$
cm$^{-2}$ (long dashed line). b) Evolution with redshift of the mean
impact parameter, for $N_{HI} = 10^{17}$ cm$^{-2}$ (dashed line) and
$N_{HI} = 1.6 \; 10^{21}$ cm$^{-2}$ (dot-dashed line).}
\label{fRbfaz1}

\end{figure}

We present in Fig.\ref{fdndlnNHIdlnRdz1} the redshift evolution of the
halo radius distribution function for a fixed column density $N_{HI} =
10^{17}$ cm$^{-2}$ (here $R$ is the radius of the dark matter halo,
i.e. without collapse). Note that massive halos are defined by a
constant radius $R=R_{cool}$. In order to take into account this
population in the picture, we also plotted in the figure the point
$(R_{cool},\pl^2n/ \pl lnN_{HI} \pl z)$ due to this contribution shown
as a filled rectangle for $z=0$ and as a small ellipse for
$z=2.5$. Thus, while the total contribution of halos defined by the
virialization constraint $\Delta=\Delta_c$ is given by the integral
over $lnN_{HI}$ of the curves shown in the figure, the total
contribution of halos defined by the cooling constraint $R=R_{cool}$
is given by the ordinate of the point $(R_{cool},\pl^2n/ \pl lnN_{HI}
\pl z)$, multiplied by a Dirac $\delta (R-R_{cool})$ function, that in
our more realistic model (VS II) is somewhat smeared out. The latter
does not appear for $z=5$ because it is negligible. This allows us to
get at once the proportion in number of both classes of objects. Thus,
one can see that at $z=0$ the ``$R_{cool}$ population'' dominates so
that most clouds associated with $N_{HI} = 10^{17}$ cm$^{-2}$ have a
dark-matter halo radius $R \sim R_{cool}=$ 120 kpc. At higher redshift
the characteristic radius decreases but one can see that at $z=2.5$
the radius distribution is still quite extended, ranging from 4 kpc up
to 120 kpc. The low radius cutoff corresponds to the transition with
the regime (1) or to the influence of the cutoff radius $R_{cut}$
(indeed to observe a high column density through a small halo one
should look close to the center where the gas density is high, however
this is not always allowed since we remove lines of sight which would
cross the galactic luminous core). It grows with time in parallel with
the ``damping'' length $R_d$ and the lower cutoff $R_c$. Note that there is
also in our present model an accumulation of clouds with radius $R_d$
(but various column densities) corresponding to population (1).

Fig.\ref{fRbfaNHI1} shows the average radius and impact parameter of
the halos associated with a given column density at $z=0, 2.5$ and
$5$. The characteristic radius and impact parameter decrease at higher
$z$ since virialized objects were smaller in the past. The panel a)
shows the same features for the average radius of Lyman-limit and
damped systems as those displayed in Fig.\ref{fvcMaNHI1} for the
circular velocity or mass, that is at first increasing with $N_{HI}$
followed by a plateau, for the same reasons as those described
previously. For Lyman forest clouds the radius $R=R_d$ is constant by
definition. As shown in panel b), the impact parameter decreases for
higher column densities, both for Lyman-limit and damped systems,
since one has to probe deeper towards the center of the clouds to
reach sufficiently high neutral hydrogen densities in order to obtain
large $N_{HI}$ (the temporary increase within the range corresponding
to Lyman-limit systems is due to the collapse of baryons which leads
to larger $b$ for a given column density because the gas density is
larger). The feature at $N_{HI} \sim 10^{20} - 10^{21}$ cm$^{-2}$ is
due to the transition between ionized shells and neutral cores, as
usual. The impact parameter is not displayed for forest clouds (regime
(1)) since it plays no role for this population and was not
specifically defined (the radius is the sole relevant scale). We note
that Bechtold et al.(1994) derive from observations at $z=1.8$ a
radius $57 < R < 400$ kpc, with a median value of $R = 130$ kpc, for
clouds with $N_{HI}
\sim 10^{14}$ cm$^{-2}$. Dinshaw et al.(1994) obtain similar
values. The average radius we obtain in Fig.\ref{fRbfaNHI1} is smaller
($\sim 21$ kpc) but as shown in Fig.\ref{fdndlnNHIdlnRdz1} the radius
distribution function is quite extended so that one indeed expects to
observe clouds with a radius up to $\sim 100$ kpc. Moreover, as discussed in Appendix, our modelling is compatible with quite elongated filamentary clouds.
The radius we refer to should be considered as the average intercept along 
the line-of-sight. However, if these very underdense clouds are filamentary, the distance at which two separate lines can hit the same cloud is much larger. So, we rather consider this offset as a possible manifestation of the
non-sphericity of the absorption features.

Fig.\ref{fRbfaz1} shows the redshift evolution of the average radius
and impact parameter for 3 column densities. One can see clearly the
decrease at high $z$ of both length scales. The decline of $b$ is
slower because at larger redshift the UV flux is lower while
characteristic densities are higher, which tends to increase the
impact parameter. We can also note that very different $N_{HI}$ have
very close mean radii and impact parameters, while the radius and
impact parameter distributions for a fixed column density are very
extended, as was shown in Fig.\ref{fdndlnNHIdlnRdz1}. Of course, this
is due to the fact that a given $N_{HI}$ can be produced by a large
range of clouds, so that very different column densities are in fact
drawn from the same population of halos.

\subsection{Metallicity}

\begin{figure}[htb]

\centerline{\epsfxsize=8 cm \epsfysize=7 cm \epsfbox{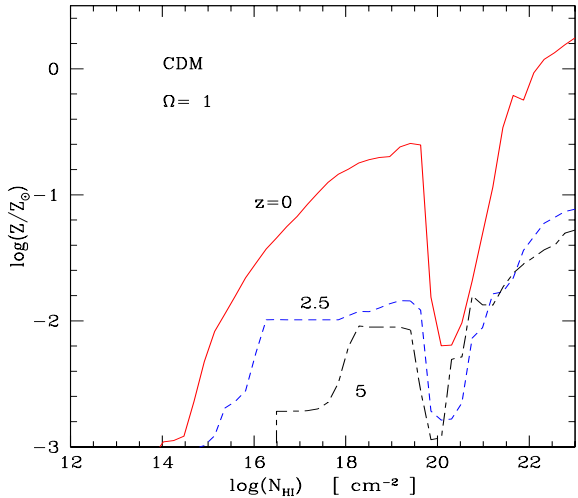}}

\caption{The average metallicity $Z_h$ of the halos associated to a 
given column density at $z=0$ (solid line), $z=2.5$ (dashed line) and
$z=5$ (dot-dashed line).}
\label{fZhaNHI1}

\end{figure}

Fig.\ref{fZhaNHI1} shows the average metallicity (that is the
abundance of oxygen or any other element mainly produced by SN II) of
the halos associated with a given column density at $z=0, 2.5$ and
$5$. We obtain this metallicity from our model for the evolution of
galaxies described in VS II. The latter is entirely consistent with
the present study of Lyman$-\alpha$ absorption lines, as it uses the
same description for the multiplicity functions of mass
condensations. It also involves the same model of star-formation which
enabled us (VS II) to get the luminosities of galaxies, as well as
their metallicity, with no new parameter. In fact, within this
framework we define three metallicities, corresponding to stars
($Z_s$), star-forming gas concentrated within the inner parts of
galaxies ($Z_c$), and diffuse gas spread over the halo ($Z_h$)
(corresponding to the two-component model of VS II). Lyman-limit
lines, which arise when the line of sight intersects the outer ionized
shells of a galactic halos, are characterized by $Z_h$, while for
damped systems, corresponding to neutral cores, we should observe a
metallicity in the range $Z_h$ to $Z_c$. We do not assign a
metallicity to low column density forest lines, corresponding to
regime (1), as they are associated with low density regions which may
have not virialized yet. Observations by Lu et al.(1998) also find
that there is a sharp drop in the metallicity of the gas (although for
$C$) from $[C/H] \simeq -2.5$ for $N_{HI} > 10^{14.5}$ cm$^{-2}$
downto $[C/H] < -3.5$ for $N_{HI} < 10^{14}$ cm$^{-2}$, at redshifts
$2.2 < z < 3.6$. Note that we predict a significant increase with $z$
of the column density corresponding with this transition. One may
expect that diffusion processes (through winds or ejection of gas
during violent mergers) would slightly lower the column density
associated with this drop and would produce a non-zero metallicity for
objects described by regime (1), which may also be enriched by a
generation of population III stars. Naturally, the metallicity
decreases at high redshift when star formation has not had enough time
to synthesize many metals. At a given redshift, the metallicity
increases with the column density, since high $N_{HI}$ implies deep
and massive potential wells. In fact, it follows closely the behaviour
of the rotational velocity as a function of the column density, see
Fig.\ref{fvcMaNHI1}, since the latter is related to the galaxy
luminosity and metallicity as shown by observations (e.g. Zaritsky et
al.1994). The sharp change for $N_{HI} \simeq 10^{20}$ cm$^{-2}$
corresponds again to the transition from ionized shells to neutral
cores.

\begin{figure}[htb]

\centerline{\epsfxsize=8 cm \epsfysize=7 cm \epsfbox{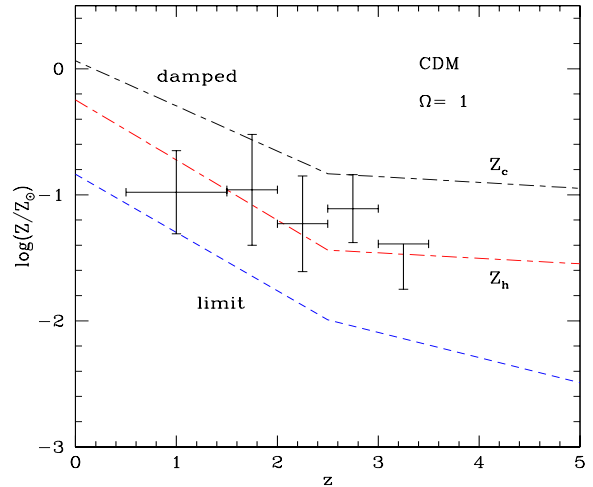}}

\caption{The redshift evolution of the average metallicity of the halos 
associated to $N_{HI}=10^{18}$ cm$^{-2}$ (dashed line) and
$N_{HI}=10^{22}$ cm$^{-2}$ (dot-dashed lines). For damped systems two
metallicities $Z_c$ and $Z_h$ are shown (see text). The data points
are from Pettini et al.(1997) for the zinc metallicity of damped
Lyman-$\alpha$ systems.}
\label{fZhaz1}

\end{figure}

Fig.\ref{fZhaz1} shows directly the redshift evolution of the
metallicity associated with $N_{HI}=10^{18}$ cm$^{-2}$ (dashed line)
and $N_{HI}=10^{22}$ cm$^{-2}$ (dot-dashed lines). For the latter case
(damped systems) we display both metallicities $Z_h$ and $Z_c$,
however the diffuse gas metallicity $Z_h$ should be the most relevant
one. We can see that we obtain very good agreement with observations
by Pettini et al.(1997). Moreover, we also obtain a large spread in
metallicities (in the same way as the velocity distribution function
was quite extended for a given column density, see
Fig.\ref{fdndlnNHIdlnvcdz1}) which can be seen in Fig.\ref{fZhaNHI1}:
although this picture only displays the average $Z_h$, the metallicity
dispersion can be estimated from the sharp variation seen near $N_{HI}
\sim 10^{21}$ cm$^{-2}$ where one probes different clouds. As was
suggested by Pettini et al.(1997) in order to explain their
observations, this is due to the fact that damped systems are drawn
from a large population of parent galactic halos which have different
star-formation histories and physical characteristics. This also
provides a good check of the validity of our model.

\subsection{Opacity}

\begin{figure}[htb]

\centerline{\epsfxsize=8 cm \epsfysize=7 cm \epsfbox{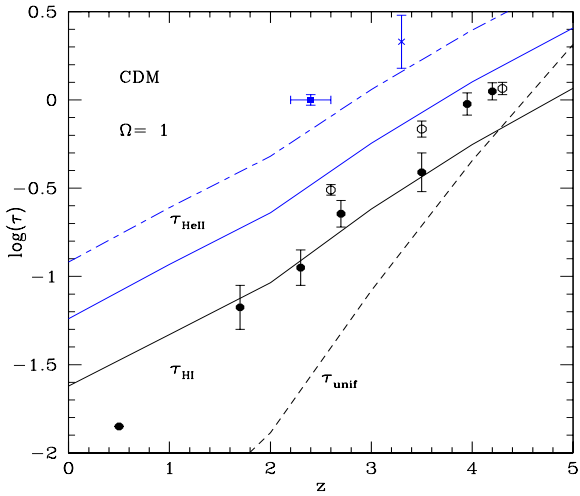}}

\caption{The evolution with redshift of the average hydrogen opacity 
$\tau_{HI}$ (lower solid line) and helium opacity $\tau_{HeII}$ (upper
solid line and dot-dashed line, see text). The low dashed line shows
the hydrogen opacity $\tau_{unif}$ which would be produced by a
uniform IGM. The data points are from Press et al.(1993) (circles),
Zuo \& Lu (1993) (filled circles) for hydrogen, and from Davidsen et
al.(1996) (filled rectangle) and Hogan et al.(1997) (cross) for
helium.}
\label{ftau1}

\end{figure}

Fig.\ref{ftau1} displays the evolution with redshift of the average
hydrogen opacity $\tau_{HI}(z)$. We can see that our result is roughly
consistent with observations. In fact, this was already implied by
Fig.\ref{fdntotodNHIdz1} and Fig.\ref{fdndz1} which showed that our
model reproduces the evolution of the column density
distribution. Indeed, the opacity is directly linked to the latter:
\beq
\tau_{HI}(z) = (1+z) \int \frac{\pl^2n}{\pl lnN_{HI} \pl z} \; 
\frac{W_{HI}}{\lambda_{\alpha}} \; dlnN_{HI}
\eeq
where $W_{HI}$ is the equivalent width. As a comparison, the dashed
line shows the opacity which would be produced by a uniform IGM:
\beq
\tau_{unif}(z) = \frac{\pi e^2 f}{m_e \nu_{\alpha}} \; \frac{n_1(z)}{H(z)}
\eeq
Thus, the clumpiness of the distribution of matter decreases the slope
of the opacity as a function of redshift, as was already noticed by
several authors (e.g. Bi \& Davidsen 1997). Indeed, from (\ref{dn2>})
we expect for a constant UV background $J_{21}$ that
\[
\tau_{HI}(z) \propto (1+z) \; \frac{dt}{dz} \; n_0(z)^{2/(2\gam-1)} \; 
 \left[ \xia^{-1}(\Delta_c) \right]^{(3-2\gam)/(2\gam-1)}
\]
If the clustering is stable, which is a good approximation for
virialized halos, we obtain:
\beq
\tau_{HI}(z) \propto (1+z)^{3(6-\gam)/(2\gam)} = (1+z)^{3.5} 
\;\;\;\; \mbox{for} \;\;\;\; \gam=1.8
\eeq
which is already very close to observations: Dobrzycki \& Bechtold
(1996) find $\tau_{HI} = 2.6 \; 10^{-3} (1+z)^{3.3}$, while
$\tau_{unif} \propto (1+z)^{4.5}$. We can also obtain the helium II
opacity provided we know the ratio $N_{HeII}/N_{HI}$. We used the
curve $N_{HeII}/N_{HI}$ calculated by Haardt \& Madau (1996) for
$J_{21} = 0.5$. In fact, as was already noticed by Miralda-Escude \&
Ostriker (1992) and Haardt \& Madau (1996), and is easily checked
numerically, most of the opacity comes from absorbers with $\tau \leq
1$ which are optically thin and verify $N_{HeII}/N_{HI} \simeq 1.8 \;
J_{912\AA}/J_{228\AA}$. Assuming that the radiation spectrum is a
power-law with index $-2$, Haardt \& Madau (1996) get $N_{HeII}/N_{HI}
\sim 30$ for these clouds. We can see from Fig.\ref{ftau1} that the
helium opacity we obtain in this way (upper solid line) is smaller
than the observations. This discrepancy may be due to a change in the
slope of the ionizing radiation spectrum. In particular, the latter
may show strong ionization edges at the HI, HeI and HeII ionization
frequencies and display a step-like profile (see Gnedin \& Ostriker
1997, Valageas \& Silk 1998) which could significantly increase the
ratio $N_{HeII}/N_{HI}$. Thus, we also show in Fig.\ref{ftau1} the
helium opacity we get with $N_{HeII}/N_{HI} = 300$ (dot-dashed
line). We note that, assuming that the $N_{HI}$ column density
distribution follows a simple power-law in column density and redshift
(chosen to be consistent with observations) and using $N_{HeII}/N_{HI}
= 100$, Zheng et al.(1998) find that at $z \sim 3$ half of the
observed opacity is accounted for by clouds with $N_{HI} > 10^{12}$
cm$^{-2}$. The helium opacity we obtain (dot-dashed line) is higher
and consistent with the data because i) we use a larger ratio
$N_{HeII}/N_{HI}$ and ii) the column density distribution we predict
in our model extends down to small objects with $N_{HI} < 10^{12}$
cm$^{-2}$, as shown by the lower cutoff $N_{HIlower}$ in
(\ref{NHI1low}) and (\ref{NHI12z}). Thus, as noticed by Zheng et
al.(1998) the large observed helium opacity strongly suggests that a
significant part of the HeII absorption is produced by small density
fluctuations which are below the observational limits for forest
clouds detected through HI absorption. This population is also a
natural prediction of our model.

\subsection{Repartition of matter between different classes of objects}

\begin{figure}[htb]

\centerline{\epsfxsize=8 cm \epsfysize=7 cm \epsfbox{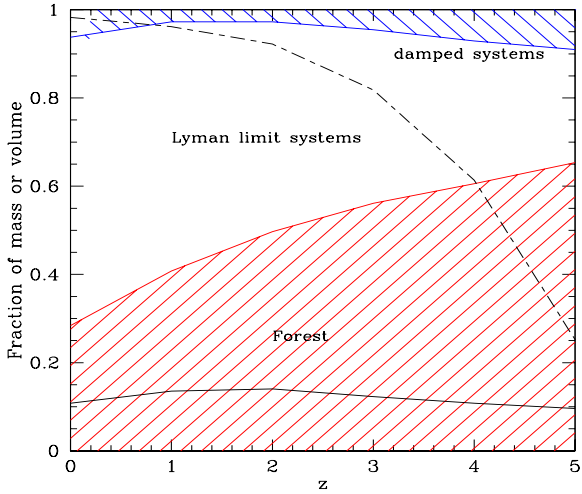}}

\caption{The evolution with redshift of the fraction of matter contained 
within Lyman-$\alpha$ clouds (for all column densities: whole lower
dashed area, and for $N_{HI}<10^{11}$ cm$^{-2}$: dashed area under the
lower solid line), Lyman limit systems (white area) and damped systems
(upper dashed area). The area above the dot-dashed line which
decreases sharply for $z>2.5$ is the volume fraction occupied by the
Lyman-$\alpha$ forest clouds with $N_{HI}>10^{11}$ cm$^{-2}$ (Lyman
limit and damped systems occupy a negligible volume fraction) while
the area under this curve is the volume fraction occupied by the
Lyman-$\alpha$ forest clouds with $N_{HI}<10^{11}$ cm$^{-2}$ and the
very low-density regions (or voids) below the threshold $N_{HIlower}$,
see (\ref{NHI1low}), which do not produce sufficiently large
absorption to be observed.}
\label{fOm1}

\end{figure}

Finally, it is interesting to evaluate the fraction of the mass of the
universe that is contained in the different populations of
Lyman-$\alpha$ clouds (it is the same proportion for baryonic and
non-baryonic matter in our model). This is displayed in Fig.\ref{fOm1}
as a function of redshift. We can see that the mass contained in Lyman
limit and damped systems (associated with galaxies) increases as time
goes on, together with structure formation via gravitational
clustering. Moreover, the mass within damped systems is always very
small, as it corresponds to the small deep cores of halos. It
increases slightly with redshift because the density of all objects is
larger at higher redshifts, which increases the relative masses of the
neutral cores as compared to the halo masses. Simultaneously, the mass
within the Lyman-$\alpha$ forest gets larger in the past when most of
the matter present in the universe is contained in low density
contrast areas which have not yet virialized (at the scale
$R_d(z)$). We must note that even at $z=0$, these clouds form $\sim
30\%$ of the mass of the universe. Thus, at any redshift an important
part of the mass of the universe is contained in small ``low density''
regions, which are not associated with galaxies or luminous
matter. This matter can only be detected through absorption lines, as
these small objects may not produce significant features in the
velocity field. Note that a sizeable proportion of this mass is
embedded within very low column density regions $N_{HI} < 10^{11}$
cm$^{-2}$. Also, the detected -neutral- hydrogen is only a small
(typically $10^{-5}$) fraction of the total baryon mass for Forest and
Lyman-limit objects. Fig.\ref{fOm1} shows the total baryon mass
fraction, including thus the dominant, but unobserved, ionised
fraction. Since we have assumed that the ratio of baryonic to total
mass is constant, the mass fractions in Fig.\ref{fOm1} also correspond to the total mass in the Lyman-$\alpha$ clouds. Note that the mass
within damped Lyman systems is small: they nevertheless stick out
prominently in the data because they correspond to totally neutral
hydrogen, a given baryon mass fraction being thus ($\sim 10^5$ times)
more efficient in producing the hydrogen absorption lines. The total
mass fraction formed by these different populations is close to unity
(the -negligible- mass fraction which we did not count corresponds to
the luminous galactic cores behind which no quasar can possibly be seen). The volume fraction occupied by Lyman-$\alpha$ forest clouds
with $N_{HI}>10^{11}$ cm$^{-2}$ is small for $z<2$ since in the highly
non-linear universe most of the volume consists of very underdense
regions. Lyman limit and damped systems occupy a negligible volume
fraction.

\subsection{Correlation function of various clouds}

\begin{figure}[htb]

\centerline{\epsfxsize=8 cm \epsfysize=7 cm \epsfbox{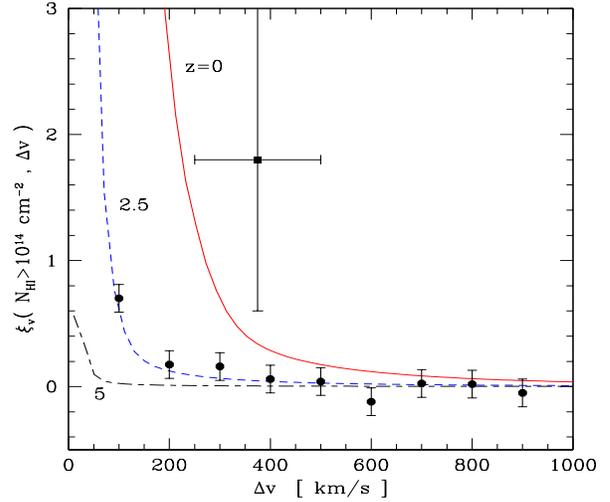}}

\caption{The two-point correlation function in velocity space for clouds 
with $N_{HI} \geq 10^{14}$ cm$^{-2}$, at the redshifts $z=0$ (solid
line), $z=2.5$ (dashed line) and $z=5$ (dot-dashed line). The data
points are from Cristiani et al.(1997) (dots) at $1.7 < z < 4$ and
from Ulmer (1996) (rectangle) at $0<z<1.3$.}
\label{fXivz1}

\end{figure}

\begin{figure}[htb]

\centerline{\epsfxsize=8 cm \epsfysize=7 cm \epsfbox{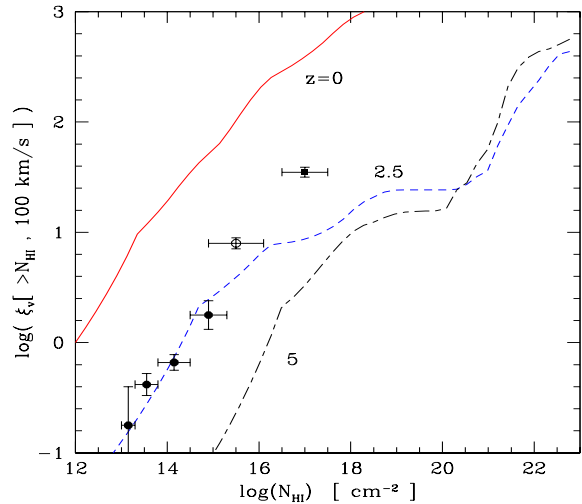}}

\caption{The dependence on the column density threshold of the two-point 
correlation function $\xi_v(>N_{HI},100$ km/s) for the redshifts $z=0$
(solid line), $z=2.5$ (dashed line) and $z=5$ (dot-dashed line). The
data points are from Cristiani et al.(1997) (disks) for $1.7 < z < 4$,
Songaila \& Cowie (1996) (circle) at $2.66 <z< 3.62$ and Petitjean \&
Bergeron (1994) (rectangle) at $z \sim 2.5$. These two latter points are
derived from CIV metal systems.}
\label{fXivNHIz1}

\end{figure}

Finally, we show in Fig.\ref{fXivz1} and Fig.\ref{fXivNHIz1} the
redshift evolution of the two-point velocity correlation function
$\xi_v$. In a fashion similar to Cristiani et al.(1997) we relate
$\xi_v$ to the spatial correlation function $\xi_{\alpha}$ of
Lyman-$\alpha$ clouds by:
\beq
\xi_v(v) = \int_{l_{min}}^{\infty} \xi_{\alpha}(r) P(v|r) dr
\eeq
with $v>0$, $r>0$. The index $\alpha$ refers to the fact that
$\xi_{\alpha}$ is linked to the matter correlation function $\xi$
through a bias factor $b^2$. We use for the conditional probability
$P(v|r)$ a gaussian of width $\sigma_v$ centered on $H r = v$, where
$H(z)$ is the Hubble constant and $\sigma_v$ a characteristic velocity
dispersion of the considered pair of clouds. The low cutoff $l_{min}$
is simply the sum of both cloud radii $l_{min} = R_1+R_2$. For small
clouds $\sigma_v \ll v$ so that $P(v|r)$ is narrowly peaked at
$r=v/H$, while for large clouds $\sigma_v \geq v$ so that the integral
is dominated by its lower cutoff, due to the divergence at small $r$
of $\xia(r)$. Thus, we write:
\beq
\xi_v(N_{HI1},N_{HI2},v) \sim \xi_{\alpha}(N_{HI1},N_{HI2},r)
\eeq
with
\beq
r = \mbox{Max} \left( R_1+R_2, \frac{v-\sigma_v}{H(z)} \right)
\eeq
for the velocity correlation function of two lines of column densities
$N_{HI1}$ and $N_{HI2}$. We estimate the pair velocity dispersion as
$\sigma_v=(v_{c1}+v_{c2})/2$, where $v_{ci}$ is the mean circular
velocity associated with a given $N_{HIi}$, which was described
previously in Fig.\ref{fvcMaNHI1} and Fig.\ref{fvcMaz1}. Next, we need
to obtain the bias parameter $b^2(N_{HI1},N_{HI2})$. Within the
framework of the scale-invariance of the many-body matter correlation
functions $\xi_p({\bf r}_1, ..., {\bf r}_p)$ which led to the mass
function (\ref{etax}) one can show that at large distances $r \gg R_1,
R_2$, in the highly non-linear regime, the bias characteristic of two
objects factorizes and is a function of the sole parameter $x$
introduced in Sect.2, see Bernardeau \& Schaeffer (1992). Thus we
write:
\beq
\xi_{\alpha}(N_{HI1},N_{HI2},r) = b(x_1) b(x_2) \xia(r)   \label{bx}
\eeq
with
\beq
x \ll 1 : \; b(x) \propto x^{(1-\omega)/2} \hspace{0.5cm} \mbox{and} 
\hspace{0.5cm} x \gg 1 : \; b(x) \propto x
\eeq
In our case, a fixed column density $N_{HI}$ may arise from many
different clouds, so as for the velocity dispersion we shall simply
use a mean $x_i$ for each $N_{HIi}$. Note that the previous relation
(\ref{bx}) breaks down for $r \sim (R_1+R_2)$, but in view of the
approximations involved (through the use of various averages) we shall
use it down to $r = (R_1+R_2)$ where it should still provide a
reasonable estimate. Finally, since observations usually refer to the
mean velocity correlation function above a given threshold in column
density, we are led to define:
\beq
\begin{array}{l} 
{\displaystyle  \xi_v(>N_{HI},v) = \left( \int_{N_{HI}}^{\infty} 
\frac{dN}{N} \eta(N) \right)^{-2} } \\ \\ {\displaystyle 
\hspace{1cm} \times \int_{N_{HI}}^{\infty} \frac{dN_1}{N_1} 
\frac{dN_2}{N_2} \eta(N_1)  \eta(N_2) \xi_v(N_1,N_2,v) }
\end{array}
\label{xiv>} 
\eeq
where $\eta(N_{HI}) = \pl^2n/ \pl lnN_{HI} \pl z$.

We compare in Fig.\ref{fXivz1} the correlation function
$\xi_v(>N_{HI},v)$ we obtain in this way to observations, for a column
density threshold $N_{HI} > 10^{14}$ cm$^{-2}$ as a function of
velocity. We can see that our predictions agree reasonably well with
observations, both at low redshift (data from Ulmer 1996) and at high
redshift (data from Cristiani et al.1997). Fig.\ref{fXivNHIz1} shows
the dependence on the column density threshold $N_{HI}$ of the
correlation function $\xi_v(>N_{HI},v)$ for the redshifts $z=0, 2.5$
and 5. We reproduce the observed increase with $N_{HI}$ of
$\xi_v(>N_{HI},v)$ over the range spanned by the data. For small
column densities, Lyman forest clouds described by the regime (1), the
correlation function measured for a fixed velocity separation
increases with $N_{HI}$ because a higher column density corresponds to
a deeper potential well, that is a higher parameter $x$, hence a
larger bias $b(N_{HI})$. Since for these low-density clouds the
circular velocity is small, as seen in Fig.\ref{fvcMaNHI1} and
Fig.\ref{fvcMaz1} (because their virial temperature $T$ is lower than
$T_0$ by definition) the spatial separation is simply constant:
$r=v/H(z)$.  Larger column densities (Lyman-limit systems) come from
deeper potential wells with a higher circular velocity so that the
bias parameter keeps increasing while the separation $r$ becomes
influenced by the velocity factor $\sigma_v$, hence decreases slowly
down to $R_1+R_2$. Thus the correlation function grows with
$N_{HI}$. As was the case for the mean radius, mass or velocity
associated to a given column density, $\xi_v(>N_{HI},v)$ reaches a
flat plateau for sufficiently large Lyman-limit systems, when
different $N_{HI}$ are drawn from the same population of parent
halos. Finally, there is also a rising part and a higher plateau for
damped systems, as in the previous studies for the velocity or
mass. We can notice that there is not a deep gap around $N_{HI} \sim
10^{20}$ cm$^{-2}$, contrary to these former cases, because the
correlation function $\xi_v(>N_{HI},v)$ involves an integral over
column densities above a given threshold, see (\ref{xiv>}), which
smooths the curves. In fact, as can be inferred from
Fig.\ref{fdntotodNHIdz1} for instance, the correlation function for
lines above $10^{19}$ cm$^{-2}$ is dominated by the contribution of
column densities around $N_{HI} \sim 10^{21}$ cm$^{-2}$.

\section{Open universe: $\Omega_0=0.3 \; , \; \Lambda=0$}

\subsection{Column density distribution}

\begin{figure}[htb]

\centerline{\epsfxsize=8 cm \epsfysize=7 cm \epsfbox{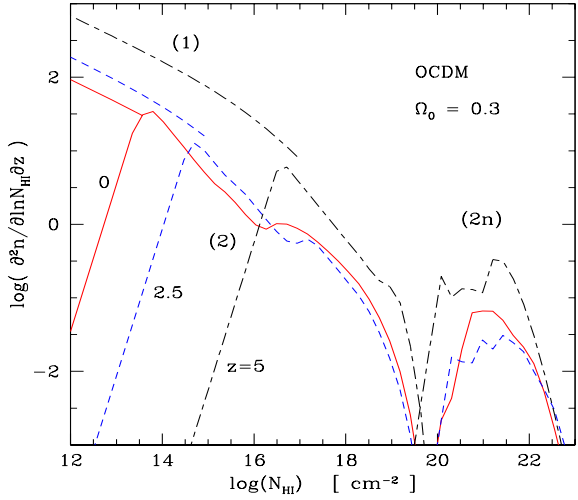}}

\caption{The contributions (1) corresponding to the Lyman-$\alpha$ 
forest (top left), (2) corresponding to the Lyman limit systems
(central part) and $(2n)$ corresponding to the damped Lyman systems
(bottom right), to the column density distribution, at $z=0$ (solid
line), $z=2.5$ (dashed line) and $z=5$ (dot-dashed line) from bottom
to top.}
\label{fdndlnNHIdz1O03}

\end{figure}

\begin{figure}[htb]

\centerline{\epsfxsize=8 cm \epsfysize=7 cm \epsfbox{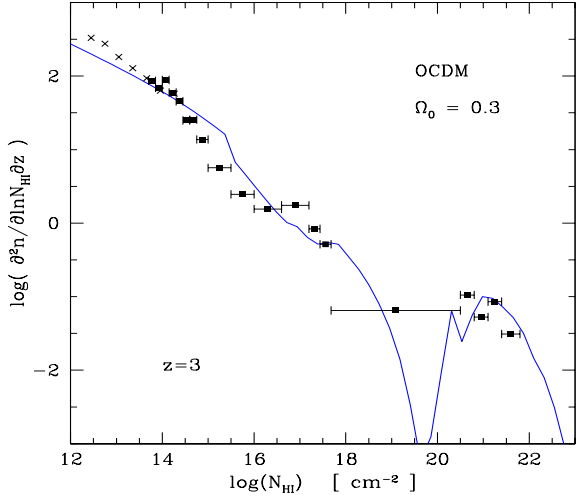}}

\caption{The total column density distribution (summed over all 
contributions) at $z=3$. The data points are from Hu et al.(1995)
(points) and from Petitjean et al.(1993) (circles).}
\label{fdntotodNHIdz1O03}

\end{figure}

\begin{figure}[htb]

\centerline{\epsfxsize=8 cm \epsfysize=7 cm \epsfbox{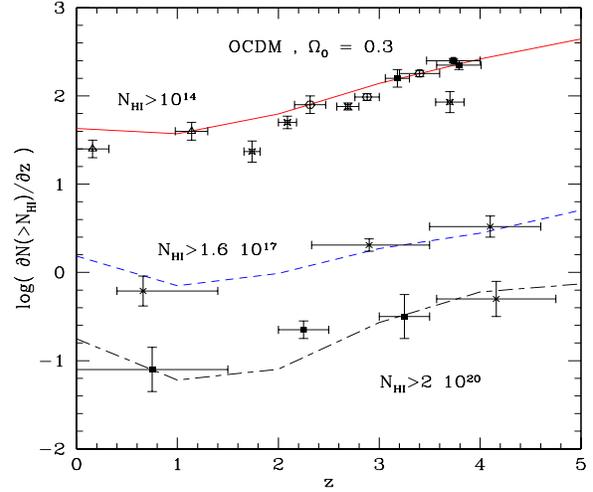}}

\caption{The evolution with redshift of the total column density 
cumulative distribution for $N_{HI} > 10^{14}$ cm$^{-2}$ (upper
curve), $N_{HI} > 1.6 \; 10^{17}$ cm$^{-2}$ (intermediary curve) and
$N_{HI} > 2 \; 10^{20}$ cm$^{-2}$ (lower curve). The data points are
from Bechtold (1994) (crosses), Bahcall et al.(1996) (circles), Giallongo
et al.(1996) (filled circles), Kim et al.(1997) (rectangles) and Lu et
al.(1996) (triangle)for $N_{HI} > 10^{14}$ cm$^{-2}$; from
Storrie-Lombardi et al.(1994) (dashed crosses) for $N_{HI} > 1.6 \;
10^{17}$ cm$^{-2}$; and from Wolfe et al.(1995) (filled circles) and
Storrie-Lombardi et al.(1995) (dashed crosses) for $N_{HI} > 2 \;
10^{20}$ cm$^{-2}$.}
\label{fdndz1O03}

\end{figure}

For a low-density universe, $\Omega_0=0.3, \; \Lambda=0$, we can
perform the same analysis. Thus, Fig.\ref{fdndlnNHIdz1O03} shows the
column density distribution at $z=0, 2.5$ and $5$, while
Fig.\ref{fdntotodNHIdz1O03} compares our predictions with observations
at $z=3$. Fig.\ref{fdndz1O03} presents the evolution with redshift of
the total column density cumulative distribution for $N_{HI}=10^{14}$
cm$^{-2}$, $N_{HI}=1.6 \; 10^{17}$ cm$^{-2}$ and $N_{HI}=2\; 10^{20}$
cm$^{-2}$. We can see that we obtain good agreement with the data,
similarly to the previous case of a critical universe. This is not
really surprising, since as we noticed earlier, provided the
model-independent properties of the multiplicity function discussed in
Sect.2 are satisfied, the main features of the column density
distribution we get, and the corresponding properties of the
Lyman-$\alpha$ clouds, come from the basic characteristics of the
physical model we built to define and recognize these absorbers. Of
course, the normalization of the power-spectrum plays some role, see
Fig.\ref{fdndlnNHIdzJOSig1} for instance, especially for damped
systems, but it is not the dominant factor for other column
densities. Thus, the values of $J_{21}(z)$ we use (see Tab.1) are
similar in both cases and consistent with observational estimates.

\subsection{Opacity}

\begin{figure}[htb]

\centerline{\epsfxsize=8 cm \epsfysize=7 cm \epsfbox{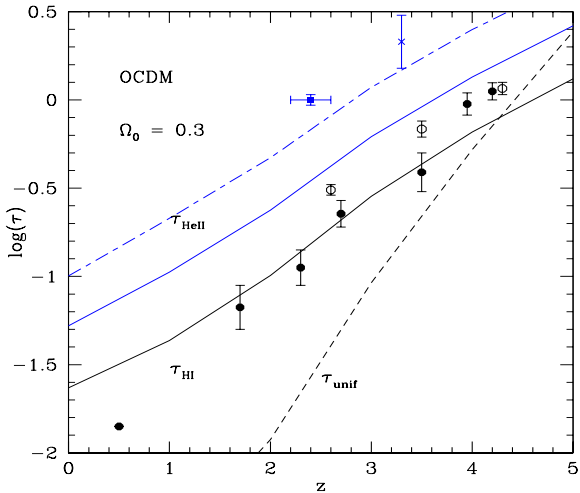}}

\caption{The evolution with redshift of the average hydrogen opacity 
$\tau_{HI}$ (lower solid line) and helium opacity $\tau_{HeII}$ (upper
solid line and dot-dashed line). The low dashed line shows the
hydrogen opacity $\tau_{unif}$ which would be produced by a uniform
IGM. The data points are from Press et al.(1993) (circles), Zuo \& Lu
(1993) (filled circles) for hydrogen, and from Davidsen et al.(1996)
(filled rectangle) and Hogan et al.(1997) (cross) for helium.}
\label{ftau1O03}

\end{figure}

Finally, Fig.\ref{ftau1O03} presents the redshift evolution of the
helium and hydrogen opacities. Of course we obtain results very close
to the case of a critical universe since the column density
distributions were already similar.

\section{Conclusion}

In this article we have developed an analytical model for the redshift
evolution of Lyman-$\alpha$ absorbers. It is based on a description of
gravitational structures which can handle objects defined by various
density contrast thresholds, both negative and positive, assuming that
the many-body correlation functions are scale-invariant. This allows
us to go beyond the usual Press-Schechter approach, which is in
principle restricted to just-virialized halos. This is a key
ingredient for modeling Lyman-$\alpha$ absorption lines since they
arise from a wide variety of objects: from underdense clouds to the
high-density neutral cores of deep potential wells. Once such a tool
to count gravitational dark matter structures is properly devised, in
a consistent way so as to keep track of all the mass in the universe
while avoiding double-counting, one still needs to specify a physical
model in order to associate with each neutral hydrogen column density
observed on a line of sight the possible objects responsible for this
feature.

Similarly to Rees (1986) we consider that large $N_{HI}$ lines come
from virialized clouds which we associate with galactic halos. The
external shells ionized by the background UV flux lead to Lyman limit
lines, while the neutral cores (protected from this UV flux by
self-shielding) correspond with damped Lyman systems. A single cloud
produces a very extended range of possible column densities, as a
function of the impact parameter of the line of sight, due to its
steep density profile ($\rho \propto r^{-\gamma}$ for our model of
spherical clouds). This has important consequences since it implies
that very different $N_{HI}$ are drawn from almost the same population
of parent halos. Thus, it leads to characteristic features (plateaus
and correspondence between various regimes) in curves such as the mean
halo rotational velocity or mass as a function of the column
density. Note that these objects are {\it not} defined, as is often
done, by the sole constant density threshold $\Delta_c \sim 177$ as we
also take into account a cooling constraint, together with the common
virialization criterium. 

Moreover, in addition to this population of clouds identified with galactic halos, we include a second class of lower density objects. These correspond to weak potential wells where baryonic density fluctuations are erased over scales of the order of the ``damping'' length $R_d$. Thus, contrary to the former class of clouds, the impact parameter plays no significant role and a given object will produce a specific column density. These halos, defined by the length scale $R_d$, cover a large range of densities, from very underdense regions
to density contrasts $\Delta \sim 40$, and correspond to the Lyman
forest lines. As discussed in Appendix, the features giving rise to these Lyman forest lines need not be real spherical objects and can be treated simply as absorbing features along the line-of-sight within our framework. Thus, although the mean intersecting length $R_{//}$ of these regions with a given line-of-sight is of order $R_d$ (which is why we also call $R_d$ the radius of these ``objects''), the actual extension of such a filament can be much larger. As a consequence, the distance at which two separate lines of sight can hit the same cloud can be much larger than $R_d$. This explains why the ``radius'' $R_d$ we obtain is much smaller than some lengths obtained from observations of close quasar pairs.

As we noticed above, this wide variety of physical objects makes it indeed necessary to use a description of gravitational structures which allows one to consider in a unified manner a large range of density contrasts and scales. For instance, one cannot study Lyman forest clouds, nor high column densities at low $z$, by looking only at just-virialized objects, while smoothing the density field over the Jeans length prevents one from modelling Lyman-limit and damped systems which correspond to very high density contrasts at similar scales.

We also have compared the redshift evolution of the column density
distribution and the hydrogen and helium opacities predicted by our
model to observations, for a critical CDM universe as well as for an
open universe $\Omega_0=0.3$. In both cases we get a good agreement
with the data, while the {\it main new parameter} is the UV flux
$J_{21}$. Indeed, the slope of the column density distribution and the
relative characteristics associated to various $N_{HI}$ are mainly
given by the physical model itself we built to identify Lyman-$\alpha$
clouds. We can also note that the amplitude of the UV flux we need is
consistent with observations and usual models of structure formation.

Next, we used the power given by such an analytical approach to study
the influence of various parameters like the normalization of the
power-spectrum $\sigma_8$, the baryonic density parameter $\Omega_b$
and the amplitude of the UV flux. Thus, we showed how their influence
varies according to the considered regime (forest, Lyman limit or
damped systems) which in principle allows one to remove for instance
the degeneracy associated with $\Omega_b^2/J_{21}$ when one is
restricted to the forest and Lyman limit contributions. Note however
that $\Omega_b$ and $\sigma_8$ are not really free parameters, since
they are chosen so as to be consistent with an earlier model of galaxy
formation. Then, we looked at detailed predictions of our model such
as the mean halo mass or radius associated with a given hydrogen
column density, as well as the velocity or radius distribution
themselves for fixed $N_{HI}$. This shows that, due to the role of the
impact parameter, such distributions are very extended so that even at
$z=2.5$, the characteristic halo rotational velocities associated for
instance with $N_{HI}=10^{17}$ cm$^{-2}$ cover a large range from
$\sim$ 10 km/s up to 500 km/s. This also means that one would need
very high-resolution numerical simulations to take into account the
contributions produced by all these clouds. Moreover, we showed that
even within the framework of the ``simple'' hierarchical structure
formation scenario the redshift evolution of quantities like the
maximum halo rotational velocity assigned to a given $N_{HI}$ displays
a qualitatively non-uniform behaviour, due to the role of
non-gravitational processes which introduce additional length or
temperature scales. Taking advantage of the fact that our model for
Lyman-$\alpha$ absorbers is part of a broader unified description of
the structures formed in the universe, including galaxies, we used the
model of galaxy formation and evolution developed earlier (VS II) to
obtain the characteristic metallicities of Lyman-limit and damped
systems. We again obtain a good agreement with observations, which
confirms the validity of our approach (both for galaxies and
Lyman-$\alpha$ clouds !).

Finally, we considered the redshift evolution of the mass and volume
fractions formed by the various populations of clouds. We note that,
contrary to some numerical studies, we managed to obtain reasonable
agreement with the observational column density distribution function
together with a UV flux $J_{21}=0.7$ at $z=2.5$ consistent with
observations using a baryonic density parameter $\Omega_b=0.04$ that
is close to nucleosynthesis bounds. As we noticed previously this
could be explained by the very wide range of parent clouds which
contribute to a given $N_{HI}$, so that it is difficult for
simulations to keep track of all objects (particularly weak potential
wells) and not to underestimate the column density distribution
function. Our formalism is also very convenient for studying the
correlation function of Lyman-$\alpha$ clouds. We again obtain results
in good agreement with observations, both for the redshift evolution
and for the dependence on column density of the amplitude of the
correlation function.

Thus, we conclude that the physical picture on which our model is
based should provide a good description of the processes at work in
the real universe, since its predictions agree with observations for
many different quantities. Moreover, it allows one to get very
detailed results and keep track of the influence of various processes,
while building a unified consistent picture of the universe. Of
course, in order to obtain a simple analytical model we had to make
some approximations: for instance we consider spherical clouds
(although we tried to correct this in a crude way in the regime where
it may make a difference: at the transition between ionized shells and
neutral cores) and we did not include the effects of star formation,
that is likely to be at the origin of the UV flux. Since the latter is
a key ingredient at all redshifts, it would be interesting to see
whether our model of galaxy formation can produce the UV flux needed
to match the observational constraints on Lyman-$\alpha$ clouds, this
will be the subject of a forthcoming article (Valageas \& Silk
1998). However, the amplitude $J_{21}$ we used is always consistent
with observational estimates, so that the results obtained in this
article are quite robust in this respect.  Eventually, the detailed
predictions of our model will have to be checked against more precise
future data, in order to narrow the range of possible physical and
cosmological parameters.

\begin{acknowledgements}
This research was supported in part by a grant from NASA.
\end{acknowledgements}

\appendix

\vspace{1cm}

{\bf APPENDIX}

\section{Objects versus density fluctuations}

In Sect.\ref{Small low-density clouds: Lyman-alpha forest} we have calculated the number of Lyman$-\alpha$ forest absorbers along the line-of-sight as if the latter were actual distinct objects of size $R_d$. We show here that there is a very simple condition for such a calculation to be equivalent to the more direct statistics of the density fluctuations along the line-of-sight. Indeed, both approaches are intimately related and there is a single condition for the latter to reduce to the former one. This also shows that we take into account all the matter along each line-of-sight (we do not restrict ourselves to the highest density peaks neglecting low density regions).

The estimate (\ref{etah}) of the number of objects of size $R$ and mass $M$ is based on the statistics of the counts in cells of size $R$. If the fluctuating density field corresponds to a density $\rho (r)$, that is to a deviation $\delta(r) = \frac {\rho(r)-\rhob} {\rhob}$, the statistics of the counts-in-cells is nothing but the statistics of $\int \delta (r) d^3r$ within the cell. On the other hand, the statistics of the density field along the line-of-sight, that we are interested in here, is the statistics of the {\it smoothed} density field over the scale $R_d$, that is the statistics of the average $\delta_d = \int \delta (r) d^3r/V_d$ where $V_d$ is the volume over which we average. The probability for the random variable $\delta_d$ to take a given value $\delta$ at a point $r$ along the line-of sight thus has the same expression as the probability to find the overdensity $\delta$ within a cell of size $R_d$, the latter being given by the statistics of the counts-in-cells (Balian \& Schaeffer 1989). It can be written:
\beq
p(\delta)d\delta = \frac {1} {\xia(R_d)} x h(x) \frac {dx} {x}
\eeq
where $x=\frac {1+\delta} {\xia(R_d)}$. The associated column density is
\beq
N_{HI} = 2 \; n_{HI} \;  R_{//} = 2 \; n_1 \; x^2 \; \xia(R_d)^2 \;R_{//}
\label {NHI 2x}
\eeq
where $R_{//}$ is the length along the line-of-sight over which the contrast $\delta$ is maintained before the density drops to a negligible value. We must recall here that at the small scales we consider here, all the matter is within very dense spots of negligible volume surrounded by a nearly empty space. This is discussed in great detail by Balian \& Schaeffer (1989) and more recently, in connection with the present work, by Valageas \& Schaeffer (1997). This equation can be usefully compared to (\ref {NHI1x}). Also, the number of occurences per redshift interval $dz$ is: 
\beq
dn =  c \frac{dt}{dz}   \frac{dz}{R_{//}} p(\delta)d\delta
 =  c \frac{dt}{dz}   \frac{dz}{R_{//}} \frac {1} {\xia(R_d)} x h(x) \frac {dx} {x} 
\label {dn}
\eeq
The column density distribution then reads
\beq
\left( \frac{\pl^2n}{\pl lnN_{HI}\pl z} \right)_1 = \frac{1}{2}
\frac{1}{R_{//}} c \frac{dt}{dz} \frac{1}{\xia(R_d)} x h(x)  
\label{dNH}
\eeq
an expression that is written for a fixed value of $N_{HI}$ and in which $x$ thus depends on $R_{//}$ through (\ref {NHI 2x}). This equation again can be usefully compared to (\ref{NHIcdd}). In the power-law regime $h(x) \propto x^{\omega -2}$ which is relevant here, it differs from the latter by a factor 
\beq
 \frac{4}{3} \frac{R_d}{R_{//}} \left( \frac{2}{3} \frac{R_d}{R_{//}} \right)^{\frac{\omega-1}{2}} 
\eeq
Provided $R_{//}$ is governed by the scale $R_d$, this factor is of order unity and our previous estimate in Sect.\ref{Small low-density clouds: Lyman-alpha forest} leads to the same results as in the present case.

Next, we can argue that $R_{//}$ actually is of order $R_d$. Indeed, since density fluctuations (and the density of typical matter condensations) increase at smaller scales (following the increase of the two-point correlation function $\xia(R)$) one expect the smallest scale at which fluctuations exist (which we called $R_d$) to dominate. This can also be seen from a simple model where we assume overdensities to be in the form of long straight cylinders of radius $R_d$ and length $L$ with $L \gg R_d$. Then, a given filament can produce different column densities as the angle $\theta$ between its main axis and the line-of-sight varies (along with the length $R_{//} \sim R_d/\sin\theta$). Indeed, a straightforward calculation along the lines of Sect.\ref{Properties of Lyman-alpha clouds} gives:
\beq
\begin{array}{l} {\displaystyle  \left( \frac{\pl^3n}{\pl lnN_{HI} 
\pl lnR_{//} \pl z} \right) \sim \frac{1}{L} c \frac{dt}{dz} \; \xia^{\;-\omega} \left( \frac{N_{HI}}{2 n_1 R_{//}} \right)^{(\omega-1)/2} }  \\  \\  {\displaystyle  \hspace{2cm} \times \; \left(1+\frac{L}{R_{//}} \right) \frac{R_d}{R_{//}} \left[ 1-\left( \frac{R_d}{R_{//}} \right)^2 \right]^{-1/2}  }
\end{array}
\label{dnRparal}
\eeq
This shows that the number of lines, for a fixed column density $N_{HI}$, is dominated by the small values of $R_{//}$: the value of the integral of (\ref{dnRparal}) over $R_{//}$ to get $\pl^2n / \pl lnN_{HI} \pl z$ is determined by the lower cutoff $R_d$. In other words, in order to achieve a given $N_{HI}$ it is more likely to look through a higher density region of size $R_d$ than to look through a lower density filament almost exactly along its main axis. This justifies our calculation and the equivalence of (\ref{dNH}) with (\ref{NHIcdd}). On the other hand, we note that the mean intersecting length $R_{//}$ obtained from (\ref{dnRparal}), that is assuming a straight cylinder, is of order $L$ (the integral now diverges at large scales). 
However, although long overdense filaments are quite likely to be present, a description in terms of a Rayleigh-Levy random walk (with direction jumps over distances of order $R_d$) rather than a long straight cylinder is certainly more accurate. In this case, the average intercept is barely twice the thickness $R_d$ and all the moments are finite, of order $R_d$. In simple words, even if the distribution of low density contrasts is filamentary, it is sufficiently irregular, with direction jumps over distances $R_d$, made of $N$ blobs of size $R_d$, but coherent over a distance $L \sim N^{1/2} R_d \gg R_d$, so as to allow us to describe the distribution of intercepts along the line-of-sight as if these features were distinct objects.
Thus, the description of the number of Lyman forest lines in terms of the fluctuating density field and our description in terms of ``objects'' of size $R_d$ are equivalent.

The main difference of such a description in terms of filamentary
clouds rather than spherical objects is the distance at which 
two different lines-of-sight can cross the same cloud. Indeed, whereas the intercept $R_{//}$ is of order $R_d$, the coherence of the absorption 
features for two neighbouring lines-of-sight will be of order $L \gg R_d$.

\end{document}